\documentclass[12pt]{article}

\usepackage{setspace}
\usepackage[utf8]{inputenc}
\usepackage[left=1in, right=1in, top=1in, lines=25]{geometry}
\usepackage[titletoc,title]{appendix}
\usepackage[svgnames]{xcolor}
\usepackage{listings}
\usepackage{changepage}
\usepackage{amsmath,amsfonts,amssymb,mathtools}
\usepackage{graphicx,float}
\usepackage{indentfirst}
\usepackage{multirow}
\usepackage{graphics}
\usepackage{caption}
\usepackage{subcaption}
\usepackage{tikz}
\usepackage{float}
\usepackage{longtable}
\usepackage[round]{natbib}
\usepackage[colorlinks=true, linkcolor=black, citecolor=black, urlcolor=black]{hyperref}
 
\allowdisplaybreaks

\usepackage{titlesec}

\titlespacing*{\section}{0pt}{1pt}{1pt}

\title{\setstretch{1.0} Multiscale Multi-Type Spatial Bayesian Analysis for High-Dimensional Data with Application to Wildfires and Migration}

\bigskip
\author
{\vspace{-1em}\textbf{Shijie Zhou} \\\vspace{-1em}
	\fontsize{8}{10}\selectfont Department of Statistics, Florida State University, 117 N. Woodward Ave., Tallahassee, Florida, U.S.A. \\\vspace{-1em}
	\fontsize{8}{10}\selectfont Email: sz20bh@fsu.edu
	\and
	\vspace{-1em}\textbf{Jonathan R. Bradley} \\\vspace{-1em}
	\fontsize{8}{10}\selectfont Department of Statistics, Florida State University, 117 N. Woodward Ave., Tallahassee, Florida, U.S.A.\\\vspace{-1em}
	\fontsize{8}{10}\selectfont Email: jrbradley@fsu.edu}\vspace{-1em}

\date{}

\begin{document}

        \maketitle

        \thispagestyle{empty}

	\begin{abstract}
		\linespread{1.1}\selectfont
		{Wildfires have significantly increased in the United States (U.S.), making certain areas harder to live in. This motivates us to jointly analyze active fires and population changes in the U.S. from July 2020 to June 2021. The available data are recorded on different scales (or spatial resolutions) and by different types of distributions (referred to as multi-type data). Moreover, wildfires are known to have feedback mechanism that creates signal-to-noise dependence. We analyze point-referenced remote sensing fire data from National Aeronautics and Space Administration (NASA) and county-level population change data provided by U.S. Census Bureau’s Population Estimates Program (PEP). We develop a multiscale multi-type spatial Bayesian model that assumes the average number of fires is zero-inflated normal, the incidence of fire as Bernoulli, and the percentage population change as normally distributed. This high-dimensional dataset makes Markov chain Monte Carlo (MCMC) implementation infeasible. We bypass MCMC by extending a recently introduced computationally efficient Bayesian framework to directly sample from the exact posterior distribution, which includes a term to model signal-to-noise dependence. Such signal-to-noise dependence is known to be present in wildfire data, but is commonly not accounted for. A simulation study is used to highlight the computational performance of our method. In our analysis, we obtained predictions of wildfire probabilities, identified several useful covariates, and found that regions with many fires were associated with population change.}
		
		
	\end{abstract}
\newpage
\pagenumbering{arabic}
\setcounter{page}{1}

	\section{Introduction}

 Wildfires incidents in the United States have notably increased over the past decades. Wildfires can last over months and affects a vast areas of lands, potentially causing people living in these highly affected areas to move \citep{McConnell21}. Not only do wildfires cause health-related problems, such as poor air quality, but they also bring negative economic impacts including property damage, lost income due to business interruptions and decreased tourism, increased healthcare costs from smoke-related illnesses, and substantial expenditures on firefighting and post-fire restoration efforts \citep{bayham22}. Consequently, it has become difficult to live in areas of the U.S. that regularly suffer from wildfires. It can be difficult to identify the effects of wildfire on migration as several studies rely on case studies (\citealp{Nawrotzki14}; \citealp{Sharygin21}; \citealp{Tinoco23}). \citet{WR21}, \citet{RW22}, and 
 \citet{McConnell24} use conditionally specified fixed and mixed effects models enforcing linear dependence in the mean when analyzing wildfire and migration. Others use summary statistics such as hypothesis tests (\citealp{Jia20}; \citealp{DeWaard23}). This motivates a joint analysis of active fires and population changes in the United States from July 2020 to June 2021 that allows for possibly nonlinear relationships. To date, no existing study has jointly modeled wildfire and migration.
 
\begin{figure}
	\centering
	\begin{subfigure}[t]{0.7\textwidth}
		\centering
		\includegraphics[width=\linewidth]{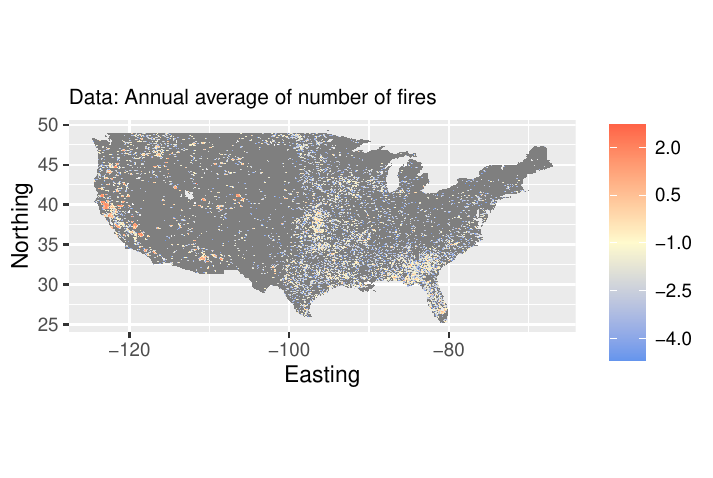} 
		\caption{}
		\label{fig:fire_data}
	\end{subfigure}
	
	\begin{subfigure}[t]{0.7\textwidth}
		\centering
		\includegraphics[width=\linewidth]{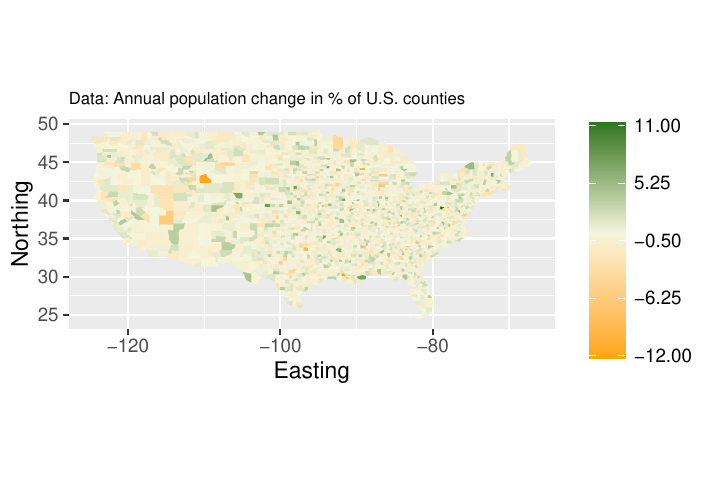} 
		\caption{}
		\label{fig:pop_data}
	\end{subfigure}
	\caption{Active fires and population changes in the United States from July 2020 to June 2021. (a) Annual average of number of fires over 80,817 point-referenced locations. The values are presented on log scale for illustration purpose only (note that we do not transform the data in our analysis), as the distribution of the data is right-skewed and presenting on its original scale makes seeing the variability of lower values difficult. The grey areas indicate that there are no fires at those locations. (b) Annual population change in percentage over 3,109 counties. Discontinuities are seen as county borders are removed for better illustration.}
	\label{fig:firenpop}
\end{figure}

 We consider remote sensing fire data from NASA \citep{nasa21}, which are available on the entire globe for a 0.1 degrees scale (i.e., $3600 \times 1800$ locations for the globe). We treat these data as point-referenced spatial data. The PEP by U.S. Census Bureau provides data of annual population change in percentage for all counties in the United States \citep{pep23}. We plot the data in Figure \ref{fig:firenpop}. The fire data in Figure \ref{fig:fire_data} is continuous with a larger (smaller) value indicating higher (lower) fire frequency at that location. Grey areas correspond to locations where there are no fires (i.e., zero-valued on the original scale of the data). In Figure \ref{fig:pop_data}, positive values (i.e., green regions) suggest increases in population for those counties while negative values (i.e., orange regions) represent decreases in population for those counties. Darker green/orange shows greater increase/decrease in population. From the plot of the data, we can observe some spatial dependence (especially in Figure \ref{fig:fire_data}); however, the relationship between these two variables is not immediate, which motivates a more sophisticated statistical analysis to identify if a bivariate relationship exists.

 It is natural to expect some cross-dependence between the fire incidents and population change. There are a number of challenges that require development in this particular dataset. First, these variables are observed on different types of spatial supports as one is point-referenced and the other is areal (i.e., regional). This is sometimes referred to as ``multiscale" in spatial statistics (see \citealp{WG04} for a standard reference). For our particular dataset, traditional multivariate spatial models may not be directly applicable because they do not allow for multiscale dependence. There are choices in the literature that allow for joint modeling of bivariate multiscale spatial data \citep{ZB24}; however, such models consider Gaussian only data and areal only multiscale case in which the data are observed at two spatially misaligned areal supports. Hence, we aim to propose a Bayesian hierarchical model (BHM) for multi-type multiscale spatial data that is mixed point-referenced and areal-referenced. Land surface temperature, rainfall, vegetation index, and elevation from NASA are included as point-referenced covariates for fire \citep{nasa21}, while median household income from the American Community Survey (ACS) is included as an areal covariate for population change \citep{acs24}. While our primary focus is on identifying potential correlation between the two multiscale responses, we are also interested in studying the effects of these covariates on their respective responses. 
 
 The second challenge in analyzing this dataset is that the fire data (denoted as $Z_1$) contains excessive zeros as about 80\% of the point-referenced locations have no fires observed (see the grey areas in Figure \ref{fig:fire_data}). Note that this does not mean these locations have missing data; instead, it shows that the annual average of number of fires at these locations are observed with a zero value. To appropriately model this characteristic of the data, we propose a conditional approach by introducing a Bernoulli variable (denoted as $Z_3$) that equals 1 if a fire is observed and 0 otherwise. We refer to this variable as the ``fire indicator variable." Then for locations where a fire is detected, we model the annual average of number of fires as normally distributed (i.e., $Z_1|Z_3=1\sim Normal$). This method allows us to distinctly model the occurrence and frequency of fires versus the observation of no fire. Such conditional approach is similar to a two-stage strategy that has been used in the past to account for zero-inflation (e.g., see \citealp{CS12} for an example in Gaussian settings; see \citealp{Wulandari23} for an example in Poisson settings). 

There are several papers that jointly model multiscale data (e.g., \citealp{GY02}; \citealp{FR05}; \citealp{Qu21}) and several papers that jointly model multi-type spatial data (e.g., \citealp{CA02}; \citealp{SH13}; \citealp{Wu15}; \citealp{Clark17}; \citealp{Jones18}; \citealp{Bradley22}; \citealp{WB24}). However, our goal is to propose a BHM for multi-type multiscale spatial data. Our approach uses a change of support (COS) method from point-referenced to county-level by adopting a basis function expansion \citep{Raim21}, and COS is achieved by integrating basis functions. Basis function expansions are particularly useful here because they imply cross-type-scale covariance and allow for nonlinear relationships that may be present in wildfire and population change (e.g., see \citealp{McConnell21} that describes that stronger relationships between these variables are present for more extreme wildfires). 

With the inclusion of the fire indicator variable that follows a Bernoulli distribution, our multi-type multiscale dataset is particularly high-dimensional with different response types. Traditionally, BHMs in such contexts are implemented using Markov chain Monte Carlo (MCMC) methods, such as the Gibbs Sampler and Metropolis-Hastings Algorithm. However, for our specific dataset, such implementations through MCMC become extremely difficult computationally due to the size of the dataset (99,205 in total) and complexity (e.g., multiscale in the multi-type settings, \citealp{Bradley22}). The high computational cost stems from the need to obtain high-dimensional dependent samples, tune the proposal densities, and ensure convergence of the samples, all of which require substantial time and effort. We bypass MCMC by extending a newly developed computationally efficient method from \citet{BC23}. Traditional Bayesian hierarchical spatial models typically involve three terms, namely, large-scale variability, small-scale variability, and fine-scale variability \citep{cressie2011statistics}. However, we consider a fourth term, which is referred to as a discrepancy term. By incorporating this new discrepancy term, \citet{BC23} shows that one can directly sampling independent replicates from the exact posterior distribution and is efficient in analyzing different types of spatial data (e.g., conditionally Gaussian, Poisson, and binomial). We further extend the use of a discrepancy term to a multivariate multiscale setting.

Wildfires are also known to produce processes that have feedback, creating signal-to-noise error (\citealp{Stavros14}; \citealp{bradley20}). In particular, wildfires release greenhouse gases, and the release of greenhouse gases contributes to higher temperatures, which in turn contributes to increases in wildfires \citep{DV19}. Hence, there is feedback between wildfires and several processes related to climate change (e.g., greenhouse gases). In observational studies such as ours, processes such as greenhouse gases are not controlled for, and consequently, the traditional additive error term implicitly includes these processes that are known to be correlated with the signal (e.g., latent mean of the annual average number of wildfires). Hence, the error associated with annual average number of wildfires realistically produces signal-to-noise correlations. Models have been developed to account for such cross-dependence \citep{bradley20,bradley23} that avoid natural identifiability issues. In particular, a fourth term is included, referred to as a discrepancy term, which produces a covariance structure that is identifiable in the likelihood. We adopt the same parametric structure used in \citet{bradley23} and \citet{BC23}, which as described, also leads to more efficient computation. Thus, not only is the consideration of a discrepancy term more computationally convenient, it is also more realistic for our data.

Motivated by the fire and population change dataset, our work proposes a multi-type multiscale BHM with a discrepancy term. Specifically, we account for the characteristic of excessive zeros in the fire data by introducing a Bernoulli variable and developing a conditional approach that is equivalent to traditional zero-inflated models in a Gaussian framework. Our extension of models with a discrepancy term to multivariate multiscale setting yields substantial computational advantage compared to MCMC by avoiding the need for tuning, convergence diagnostics, etc. The remainder of this paper will be organized as follows. In Section 2, we will review the Diaconis-Ylvisaker (DY) distribution and generalized conjugate multivariate (GCM) distribution, which are essential in the development of our method. In Section 3, we specify our multi-type multiscale BHM. Section 4 presents a comprehensive simulation study in which we compare our method with the traditional MCMC method in terms of predictive accuracy and computational efficiency. We show the real analysis results for the U.S. fire and population change dataset in Section 5. A discussion is followed in Section 6.

	\section{Reviews}

 We make use of the DY distribution and recent development in the multivariate context. To aid the reader, we provide brief reviews of the DY distribution (Section 2.1) and the GCM distribution (Section 2.2).

\subsection{Exponential Family and Diaconis-Ylvisaker Distribution}

Suppose data $Z$ follows a distribution that is a member from the exponential family, 
\begin{align}
    \label{eq:expofam}
    f(Z|Y)=\exp\{ZY-b\psi(Y)+c(Z)\},
\end{align}
where $Z$ denotes the data, $Y$ represents the natural parameter, $b$ is a scalar that may be known, $\psi(\cdot)$ is the known unit log-partition function, and $c(\cdot)$ is a known function. Different forms of $b$ and $\psi(\cdot)$ correspond to different distributions in the exponential family. In our motivating dataset, we focus on Gaussian and Bernoulli distributed data. Specifically, $\psi(Y)\equiv \psi_G(Y)=Y^2$, $b\equiv b_G = \frac{1}{2\sigma^2}$ when the data is Gaussian (denote with the subscript ``G") for $Z\in \mathbb{R}$, $Y\in \mathbb{R}$, and $\sigma^2>0$, while $\psi(Y)\equiv \psi_B(Y)=\log\{1+\exp(Y)\}$, $b\equiv b_B=1$ when the data is Bernoulli (denote with the subscript ``B") for $Z\in\{0,1\}$ and $Y\in \mathbb{R}$. \citet{DY79} developed the conjugate distribution in these cases, which is referred to as the DY distribution given by,
\begin{align}
    \label{eq:dy}
    f(Y|\alpha,\kappa)\propto\exp\{\alpha Y-\kappa\psi(Y)\},
\end{align}
where $\kappa>0$, $Y\in\mathbb{R}$, $\frac{\alpha}{\kappa}\in\mathbb{R}$ for Gaussian data, and $\kappa>\alpha$, $Y\in\mathbb{R}$, $\alpha>0$ for Bernoulli data. We denote it as $Y|\alpha,\kappa\sim DY(\alpha,\kappa;\psi)$. This particular form of the DY distribution in (\ref{eq:dy}) is conjugate with the exponential families in 
(\ref{eq:expofam}) since
\begin{align*}
    f(Y|Z,\alpha,\kappa)\propto f(Z|Y)f(Y|\alpha,\kappa)\propto\exp\{(Z+\alpha)Y-(b+\kappa)\psi(Y)\}.
\end{align*}
Therefore, $Y|Z,\alpha,\kappa\sim DY(Z+\alpha,b+\kappa;\psi)$.

\subsection{Generalized Conjugate Multivariate Distribution}

\citet{BC23} defined the generalized conjugate multivariate (GCM) distribution based on the transformation,
\begin{align}
    \label{eq:transformation}
    \mathbf{y}=\boldsymbol{\mu}+\boldsymbol{V}\boldsymbol{D}(\boldsymbol{\theta})\boldsymbol{w},
\end{align}
where $\mathbf{y}$ is an $n$-dimensional random vector, $\boldsymbol{\mu}$ is an unknown $n$-dimensional location vector, $\boldsymbol{V}$ is an $n\times n$ invertible covariance parameter matrix, $\boldsymbol{D}:\Omega\rightarrow \mathbb{R}^n\times \mathbb{R}^n$ is a known matrix valued function such that the inverse $\boldsymbol{D}(\boldsymbol{\theta})^{-1}$ exists for any hyperparameters $\boldsymbol{\theta}\in\Omega$, and $\boldsymbol{w}$ is a random vector that contains $n$ independent and not identically distributed DY random variables. We assume $\boldsymbol{\theta}$ follows a distribution with proper density $\pi(\boldsymbol{\theta})$. In our multi-type case with Gaussian and Bernoulli data, we can write $\boldsymbol{w}=(\boldsymbol{w}_G',\boldsymbol{w}_B')'$, where $\boldsymbol{w}_G=(w_{G,1},...,w_{G,n_G})'$ and $\boldsymbol{w}_B=(w_{B,1},...,w_{B,n_B})'$. Note that $n_G$ and $n_B$ represent the numbers of observations of Gaussian and Bernoulli distributed data, respectively, such that $n_G+n_B=n$. Then $w_{G,i}\sim DY(\alpha_{G,i},\kappa_{G,i};\psi_G)$ and $w_{B,j}\sim DY(\alpha_{B,j},\kappa_{B,j};\psi_B)$, where $\kappa_{G,i}>0$, $\frac{\alpha_{G,i}}{\kappa_{G,i}}\in \mathbb{R}$, $\kappa_{B,j}>\alpha_{B,j}$, $\alpha_{B,j}>0$, and the known unit log partition functions $\psi_G(w)=w^2$ and $\psi_B(w)=\log\{1+\exp(w)\}$ as defined in Section 2.1. We let $\boldsymbol{\alpha}=(\alpha_{G,1},...,\alpha_{G,n_G},\alpha_{B,1},...,\alpha_{B,n_B})'$, $\boldsymbol{\kappa}=(\kappa_{G,1},...,\kappa_{G,n_G},\kappa_{B,1},...,\kappa_{B,n_B})'$, and $\boldsymbol{\psi}(\boldsymbol{w})=(\boldsymbol{\psi}_G(\boldsymbol{w}_G)',\boldsymbol{\psi}_B(\boldsymbol{w}_B)')'$ where $\boldsymbol{\psi}_G(\boldsymbol{w}_G)=(\psi_G(w_{G,1}),...,\psi_G(w_{G,n_G}))'$ and $\boldsymbol{\psi}_B(\boldsymbol{w}_B)=(\psi_B(w_{B,1}),...,\psi_B(w_{B,n_B}))'$. The PDF of $\mathbf{y}$ is then given by,
\begin{align}
    \label{eq:gcm}
f(\mathbf{y}|\boldsymbol{\mu},\boldsymbol{V},\boldsymbol{\alpha},\boldsymbol{\kappa})=\int_{\Omega}\pi(\boldsymbol{\theta})N\exp\left[\boldsymbol{\alpha}'\boldsymbol{D}(\boldsymbol{\theta})^{-1}\boldsymbol{V}^{-1}(\mathbf{y}-\boldsymbol{\mu})-\boldsymbol{\kappa}'\boldsymbol{\psi}\left\{\boldsymbol{D}(\boldsymbol{\theta})^{-1}\boldsymbol{V}^{-1}(\mathbf{y}-\boldsymbol{\mu})\right\}\right]d\boldsymbol{\theta},
\end{align}
where $N$ is a normalizing constant. We denote (\ref{eq:gcm}) as $\mathbf{y}\sim GCM(\boldsymbol{\mu},\boldsymbol{V},\boldsymbol{\alpha},\boldsymbol{\kappa};\boldsymbol{\psi})$. It is also developed by \citet{BC23} that if $\mathbf{y}=(\mathbf{y}_1',\mathbf{y}_2')'\sim GCM(\boldsymbol{\mu},\boldsymbol{V},\boldsymbol{\alpha},\boldsymbol{\kappa};\boldsymbol{\psi})$, where $\mathbf{y}_1$ is $r$-dimensional and $\mathbf{y}_2$ is $(n-r)$-dimensional, and $\boldsymbol{V}^{-1}=(\boldsymbol{H},\boldsymbol{Q})$, where $\boldsymbol{H}$ and $\boldsymbol{Q}$ are $n\times r$ and $n\times (n-r)$ submatrices, then $\mathbf{y}_1|\mathbf{y}_2$ follows a conditional GCM (cGCM) such that
\begin{align}
\label{eq:cGCM}
f(\mathbf{y}_1|\mathbf{y}_2,\boldsymbol{\mu},\boldsymbol{V},\boldsymbol{\alpha},\boldsymbol{\kappa}) \propto \int_{\Omega}\frac{\pi(\boldsymbol{\theta})}{\det\{\boldsymbol{D}(\boldsymbol{\theta})\}}\exp\left[\boldsymbol{\alpha}'\{\boldsymbol{D}(\boldsymbol{\theta})^{-1}\boldsymbol{H}\mathbf{y}_1-\boldsymbol{\mu}^*\}-\boldsymbol{\kappa}'\boldsymbol{\psi}\{\boldsymbol{D}(\boldsymbol{\theta})^{-1}\boldsymbol{H}\mathbf{y}_1-\boldsymbol{\mu}^*\}\right]d\boldsymbol{\theta},
\end{align}
where $\boldsymbol{\mu}^*=\boldsymbol{D}(\boldsymbol{\theta})^{-1}\boldsymbol{V}^{-1}\boldsymbol{\mu}-\boldsymbol{D}(\boldsymbol{\theta})^{-1}\boldsymbol{Q}\mathbf{y}_2$. We denote (\ref{eq:cGCM}) as $\mathbf{y}_1|\mathbf{y}_2\sim cGCM(\boldsymbol{\mu}^*,\boldsymbol{H},\boldsymbol{\alpha},\boldsymbol{\kappa};\boldsymbol{\psi})$. \citet{BC23} show that the posterior distribution associated with spatial generalized linear mixed effects models follow a cGCM distribution. Outside of certain settings (e.g., Gaussian), it is not known currently how to sample from cGCM. However, one can augment this model such that the posterior is GCM, which we know how to simulate from directly using Equation (\ref{eq:transformation}). We adopt this strategy and extend it to the multiscale multi-type spatial setting.
	
	\section{Methodologies}

    In Section 3.1, we state our model. Then in Section 3.2, we provide theoretical development that aid in computation and interpretation of a particular term in our model, referred to as the discrepancy term.

    \subsection{Model}
	
	We let $\{Z_1(\boldsymbol{s}_1),...,Z_1(\boldsymbol{s}_{n_1})\}$ and $\{Z_2(A_1),...,Z_2(A_{n_2})\}$ be the observed annual average number of fires and annual percentage population change, respectively. Here, each $\boldsymbol{s}_i$ represents a point-referenced location in the contiguous U.S., while each $A_j\subset \mathbb{R}^2$ represents a U.S. county or county-equivalent. A large number of values in $\{Z_1(\boldsymbol{s}_i)\}$ are zero, and as such, we adopt an approach that incorporates an indicator variable similar to \citet{CS12}. Specifically, we introduce a Bernoulli-distributed fire indicator variable $Z_3(\boldsymbol{s}_i)$ such that $Z_3(\boldsymbol{s}_i)=1$ if there are some fires observed at location $\boldsymbol{s}_i$ (i.e., $Z_1(\boldsymbol{s}_i)>0$) and $Z_3(\boldsymbol{s}_i)=0$ if there is no fire observed (i.e., $Z_1(\boldsymbol{s}_i)=0$). We assume $Z_1(\boldsymbol{s}_i)|Z_3(\boldsymbol{s}_i)=1$ follows a Gaussian distribution, and $Z_2(A_j)$ is also assumed to be Gaussian distributed. In particular, the data models can be written as follows,
    \begin{align}
	\label{eq:datamodel}	Z_1(\boldsymbol{s}_i)&|Z_3(\boldsymbol{s}_i)=1,\boldsymbol{\beta}_1,\boldsymbol{\eta}_1,\boldsymbol{\eta}_2,\xi_1(\boldsymbol{s}_i),\delta_1(\boldsymbol{s}_i),\sigma_{1i}^2 \sim Normal(Y_1(\boldsymbol{s}_i),\sigma_{1i}^2) \nonumber \\
	&\quad where \,\,Y_1(\boldsymbol{s}_i)=\boldsymbol{x}_1(\boldsymbol{s}_i)'\boldsymbol{\beta}_1+\boldsymbol{g}_1(\boldsymbol{s}_i)'\boldsymbol{\eta}_1+\boldsymbol{g}_1(\boldsymbol{s}_i)'\boldsymbol{\eta}_2+(\xi_1(\boldsymbol{s}_i)-\delta_1(\boldsymbol{s}_i))  \nonumber \\
	Z_2(A_j)&|\boldsymbol{\beta}_2,\boldsymbol{\eta}_1,\boldsymbol{\eta}_3,\xi_2(A_j),\delta_2(A_j),\sigma_{2j}^2\sim Normal(Y_2(A_j),\sigma_{2j}^2)  \\
	&\quad where\,\,Y_2(A_j)=\boldsymbol{x}_2(A_j)'\boldsymbol{\beta}_2+\boldsymbol{g}_2(A_j)'\boldsymbol{\eta}_1+\boldsymbol{g}_2(A_j)'\boldsymbol{\eta}_3+(\xi_2(A_j)-\delta_2(A_j)) \nonumber\\
	Z_3(\boldsymbol{s}_k)&|\boldsymbol{\beta}_3,\boldsymbol{\eta}_1,\xi_3(\boldsymbol{s}_k),\delta_3(\boldsymbol{s}_k)\sim Bernoulli\left\{\frac{\exp\left(Y_3(\boldsymbol{s}_k)\right)}{1+\exp\left(Y_3(\boldsymbol{s}_k)\right)}\right\} \nonumber \\
	\quad &\quad where\,\,Y_3(\boldsymbol{s}_k)=\boldsymbol{x}_3(\boldsymbol{s}_k)'\boldsymbol{\beta}_3+\boldsymbol{g}_3(\boldsymbol{s}_k)'\boldsymbol{\eta}_1+(\xi_3(\boldsymbol{s}_k)-\delta_3(\boldsymbol{s}_k)) \nonumber,
\end{align}
 for $i=1,...,n_1^*$, $j=1,...,n_2$, and $k=1,...,n_1$. Note that $n_1^*<n_1$ where $n_1^*$ is the number of point-referenced locations with nonzero annual average number of fires and $n_1$ is the number of all point-referenced locations. For $m=1,2,3$, $\boldsymbol{x}_m$ are $p_m$-dimensional column vectors of known covariates, respectively, and $\boldsymbol{\beta}_m$ are the corresponding $p_m$-dimensional unknown regression coefficients. We let $\boldsymbol{g}_m$ to be prespecified $r$-dimensional spatial basis functions. We allow for multiple scales through $\boldsymbol{g}_m$. In particular, we assume Gaussian radial basis functions for $\boldsymbol{g}_m(\boldsymbol{s})$ and perform spatial COS through integration such that,
\begin{align*}
	\boldsymbol{g}_m(A_j)=\frac{1}{|A_j|}\int_{A_j} \boldsymbol{g}_m(\boldsymbol{s})\,d\boldsymbol{s},
\end{align*}
where $|A_j|$ denotes the area of $A_j$. The associated $\boldsymbol{\eta}_m$ are $r$-dimensional vectors of random effects. The random effect $\boldsymbol{\eta}_1$ allows for cross-type covariances (e.g., $cov(Y_1(\boldsymbol{s}_i),Y_2(A_j))=\boldsymbol{g}_1(\boldsymbol{s}_i)'cov(\boldsymbol{\eta}_1)\boldsymbol{g}_2(A_j)\neq 0$). When examining Figure \ref{fig:fire_data}, the presence of a fire (i.e., $Z_3$) displays a functionally simplistic pattern: high incidences of fire in the far west coast, followed by the mid-west and southeast, and some fine-scale variability throughout the spatial domain. Consequently, we assume fewer basis functions (than $Y_1$ and $Y_2$) to define $Y_3$ to model the simplistic small-scale variability, and include a fine-scale variability term $\xi_3$. The presence of $\boldsymbol{\eta}_2$ and $\boldsymbol{\eta}_3$, and $\boldsymbol{g}_i\neq \boldsymbol{g}_j$ for $i\neq j$ implies that each response has a different marginal spatial covariance. The $\xi_m$ are the terms for uncorrelated (or weakly correlated) variability that is typically assumed to be normally distributed with mean zero and some variance. The aforementioned components of the model are common specifications in spatial statistics and are traditionally referred to as large-scale variability (i.e., $\boldsymbol{x}_m'\boldsymbol{\beta}_m$), small-scale variability (i.e., $\boldsymbol{g}_m'\boldsymbol{\eta}_m$), and fine-scale variability (i.e., $\xi_m$), respectively \citep{cressie2011statistics}. However, availing of results from \citet{BC23}, we introduce an extra discrepancy term (i.e., $\delta_m$), which is partially motivated by modeling signal-to-noise covariance between the signal $Y_m$ and the noise $\delta_m$.

 Let $n\equiv n_1^*+n_2+n_1$ and $p\equiv p_1+p_2+p_3$. We organize the latent random variables in (\ref{eq:datamodel}) into an $n$-dimensional vector, $\mathbf{y}=(Y_1(\boldsymbol{s}_1),...,Y_1(\boldsymbol{s}_{n_1^*}),Y_2(A_1),...,Y_2(A_{n_2}),Y_3(\boldsymbol{s}_1),...,Y_3(\boldsymbol{s}_{n_1}))'$. We can express $\mathbf{y}$ as follows,
\begin{equation}
	\begin{aligned}
		\mathbf{y}&=\begin{pmatrix}
			\boldsymbol{X}_1 & \boldsymbol{0} & \boldsymbol{0} \\
			\boldsymbol{0} & \boldsymbol{X}_2 & \boldsymbol{0} \\
			\boldsymbol{0} & \boldsymbol{0} & \boldsymbol{X}_3
		\end{pmatrix}\begin{pmatrix}
			\boldsymbol{\beta}_1 \\
			\boldsymbol{\beta}_2 \\
			\boldsymbol{\beta}_3 
		\end{pmatrix}+\begin{pmatrix}
			\boldsymbol{G}_1 & \boldsymbol{G}_1 & \boldsymbol{0} \\
			\boldsymbol{G}_2 & \boldsymbol{0} & \boldsymbol{G}_2 \\
			\boldsymbol{G}_3 & \boldsymbol{0} & \boldsymbol{0}
		\end{pmatrix}\begin{pmatrix}
			\boldsymbol{\eta}_1 \\
			\boldsymbol{\eta}_2 \\
			\boldsymbol{\eta}_3 
		\end{pmatrix}+\left\{\begin{pmatrix}
			\boldsymbol{\xi}_1 \\
			\boldsymbol{\xi}_2 \\
			\boldsymbol{\xi}_3 
		\end{pmatrix}-\begin{pmatrix}
			\boldsymbol{\delta}_1 \\
			\boldsymbol{\delta}_2 \\
			\boldsymbol{\delta}_3 
		\end{pmatrix}\right\}\\
		&\equiv \boldsymbol{X\beta}+\boldsymbol{G\eta}+(\boldsymbol{\xi}-\boldsymbol{\delta}_y),
	\end{aligned}
	\label{eq:process}
\end{equation}
where the concatenated covariates matrix $\boldsymbol{X}$ has dimension $n\times p$ with the $n_1^*\times p_1$ matrix $\boldsymbol{X}_1=(\boldsymbol{x}_1(\boldsymbol{s}_1)',...,\boldsymbol{x}_1(\boldsymbol{s}_{n_1^*})')'$, $n_2\times p_2$ matrix $\boldsymbol{X}_2=(\boldsymbol{x}_2(A_1)',...,\boldsymbol{x}_2(A_{n_2})')'$, and $n_1\times p_3$ matrix $\boldsymbol{X}_3=(\boldsymbol{x}_3(\boldsymbol{s}_1)',...,\boldsymbol{x}_3(\boldsymbol{s}_{n_1})')'$ on the diagonal; $\boldsymbol{G}$ is the concatenated $n\times 3r$ basis functions matrix, where $\boldsymbol{G}_1=(\boldsymbol{g}_1(\boldsymbol{s}_1)',...,\boldsymbol{g}_1(\boldsymbol{s}_{n_1^*})')'$, $\boldsymbol{G}_2=(\boldsymbol{g}_2(A_1)',...,\boldsymbol{g}_2(A_{n_2})')'$, and $\boldsymbol{G}_3=(\boldsymbol{g}_3(\boldsymbol{s}_1)',...,\boldsymbol{g}_3(\boldsymbol{s}_{n_1})')'$; the $p$-dimensional $\boldsymbol{\beta}$ and $3r$-dimensional $\boldsymbol{\eta}$ are the stacked fixed and random effects, respectively; the $n$-dimensional fine-scale variability $\boldsymbol{\xi}=(\boldsymbol{\xi}'_1,\boldsymbol{\xi}'_2,\boldsymbol{\xi}'_3)'$ where $\boldsymbol{\xi}_1=(\xi_1(\boldsymbol{s}_1),...,\xi_1(\boldsymbol{s}_{n_1^*}))'$, $\boldsymbol{\xi}_2=(\xi_2(A_1),...,\xi_2(A_{n_2}))'$, and $\boldsymbol{\xi}_3=(\xi_3(\boldsymbol{s}_1),...,\xi_3(\boldsymbol{s}_{n_1}))'$; similarly, the $n$-dimensional discrepancy term $\boldsymbol{\delta}_y=(\boldsymbol{\delta}'_1,\boldsymbol{\delta}'_2,\boldsymbol{\delta}'_3)'$ where $\boldsymbol{\delta}_1=(\delta_1(\boldsymbol{s}_1),...,\delta_1(\boldsymbol{s}_{n_1^*}))'$, $\boldsymbol{\delta}_2=(\delta_2(A_1),...,\delta_2(A_{n_2}))'$, and $\boldsymbol{\delta}_3=(\delta_3(\boldsymbol{s}_1),...,\delta_3(\boldsymbol{s}_{n_1}))'$.

We assume that $\boldsymbol{\beta}$ has a Gaussian prior with $p$-dimensional location vector $\boldsymbol{D}_{\beta}(\boldsymbol{\theta})\boldsymbol{\delta}_{\beta}$ and $p\times p$ covariance matrix $\boldsymbol{D}_{\beta}(\boldsymbol{\theta})\boldsymbol{D}_{\beta}(\boldsymbol{\theta})'$. Similarly, $\boldsymbol{\eta}$ is also assumed a Gaussian prior with $3r$-dimensional location vector $\boldsymbol{D}_{\eta}(\boldsymbol{\theta})\boldsymbol{\delta}_{\eta}$ and $3r\times3r$ covariance matrix $\boldsymbol{D}_{\eta}(\boldsymbol{\theta})\boldsymbol{D}_{\eta}(\boldsymbol{\theta})'$. Here, $\boldsymbol{\theta}$ represents a vector of hyperparameters, $\boldsymbol{D}_{\beta}(\boldsymbol{\theta}):\Omega\rightarrow\mathbb{R}^{p}\times\mathbb{R}^{p}$, and $\boldsymbol{D}_{\eta}(\boldsymbol{\theta}):\Omega\rightarrow\mathbb{R}^{3r}\times\mathbb{R}^{3r}$. In traditional spatial BHM without the discrepancy term, the fine-scale variability $\boldsymbol{\xi}$ is often also assumed Gaussian \citep{cressie2011statistics}. However, we assume it follows a cGCM that is close to a Gaussian distribution, which is similar to the distributional assumption used in \citet{BC23}. Specifically, $\boldsymbol{\xi}$ is proportional to $cGCM(\boldsymbol{\alpha}_{\xi},\boldsymbol{\kappa}_{\xi},\boldsymbol{\delta}^*_{\xi},\boldsymbol{H}_{\xi},\pi_{\xi},\boldsymbol{D}_{\xi};\boldsymbol{\psi}_{\xi})$, where the $2n$-dimensional $\boldsymbol{\alpha}_{\xi}=(\boldsymbol{0}_{1,n_1^*},\boldsymbol{0}_{1,n_2},\alpha_{\xi}\boldsymbol{1}_{1,n_1},\boldsymbol{0}_{1,n})'$, the $2n$-dimensional $\boldsymbol{\kappa}_{\xi}=(\boldsymbol{0}_{1,n_1^*},\boldsymbol{0}_{1,n_2},2\alpha_{\xi}\boldsymbol{1}_{1,n_1},\frac{1}{2}\boldsymbol{1}_{1,n})'$, the $2n$-dimensional $\boldsymbol{\delta}^*_{\xi}=(\boldsymbol{\delta}'-\boldsymbol{\beta}'\boldsymbol{X}'-\boldsymbol{\eta}'\boldsymbol{G}',\boldsymbol{\delta}_{\xi}')'$, and the $2n\times n$ matrix $\boldsymbol{H}_{\xi}=(\sigma_{\xi}\boldsymbol{I}_n,\boldsymbol{I}_n)'$. Note that the scalars $\alpha_{\xi}>0$ and $\sigma^2_{\xi}>0$, $\boldsymbol{0}_{i,j}$ and $\boldsymbol{1}_{i,j}$ are $(i\times j)$-dimensional vectors/matrices of zeros and ones, respectively, and $\boldsymbol{\delta}_{\xi}$ is an $n$-dimensional real vector. We further let $\pi_{\xi}$ to be an indicator function such that $\pi_{\xi}(\theta)=I(\theta=\sigma^2_{\xi})$, $\boldsymbol{D}_{\xi}=\sigma_{\xi}\boldsymbol{I}_{2n}$ where $\boldsymbol{I}_k$ is a $k \times k$ identity matrix, and $\boldsymbol{\psi}_{\xi}(\boldsymbol{h}_{\xi})=(\psi_G(h_{1,1}),...,\psi_G(h_{1,n_1^*}),\psi_G(h_{2,1}),...,\psi_G(h_{2,n_2}),\psi_B(h_{3,1}),...,\psi_B(h_{3,n_1}),\psi_G(h^*_{1,1}),$\newline$...,\psi_G(h^*_{1,n_1^*}),\psi_G(h^*_{2,1}),...,\psi_G(h^*_{2,n_2}),\psi_G(h^*_{3,1}),...,\psi_G(h^*_{3,n_1}))'$ for any $\boldsymbol{h}_{\xi}=(h_{1,1},...,h_{1,n_1^*},h_{2,1},$\newline$...,h_{2,n_2},h_{3,1},...,h_{3,n_1},h^*_{1,1},...,h^*_{1,n_1^*},h^*_{2,1},...,h^*_{2,n_2},h^*_{3,1},...,h^*_{3,n_1})'\in \mathbb{R}^{2n}$.

We store the $n$-dimensional $\boldsymbol{\delta}_y$, $p$-dimensional $\boldsymbol{\delta}_{\beta}$, $3r$-dimensional $\boldsymbol{\delta}_{\eta}$, and $n$-dimensional $\boldsymbol{\delta}_{\xi}$ in a single vector and write $\boldsymbol{\delta}=(\boldsymbol{\delta}_y',\boldsymbol{\delta}_{\beta}',\boldsymbol{\delta}_{\eta}',\boldsymbol{\delta}_{\xi}')'$. We let $\boldsymbol{\delta}=-\boldsymbol{D}(\boldsymbol{\theta})^{-1}\boldsymbol{Qq}$ and assume an improper prior on $\boldsymbol{q}$ (i.e.,$f(\boldsymbol{q})\propto1$), then the posterior distribution of $(\boldsymbol{\xi}',\boldsymbol{\beta}',\boldsymbol{\eta}',\boldsymbol{q}')'$ would follow a GCM distribution (see Section 3.2), which we can directly sample from without the need for MCMC \citep{BC23}. Here, $\boldsymbol{D}(\boldsymbol{\theta})^{-1}$ is a block diagonal matrix such that $\boldsymbol{D}(\boldsymbol{\theta})^{-1}=blkdiag(\boldsymbol{I}_n,\boldsymbol{D}_{\beta}(\boldsymbol{\theta})^{-1},\boldsymbol{D}_{\eta}(\boldsymbol{\theta})^{-1},\frac{1}{\sigma_\xi}\boldsymbol{I}_n)$, and the $(2n+p+3r)\times n$ matrix $\boldsymbol{Q}$ contains the eigenvectors of the orthogonal complement of $\boldsymbol{H}$, where 
\begin{align}
\label{eq:Hmatrix}
    \boldsymbol{H}=\begin{pmatrix}
\boldsymbol{I}_n & \boldsymbol{X} & \boldsymbol{G}\\
\boldsymbol{0}_{p,n} & \boldsymbol{I}_p & \boldsymbol{0}_{p,3r}\\
\boldsymbol{0}_{3r,n} & \boldsymbol{0}_{3r,p} & \boldsymbol{I}_{3r} \\
\boldsymbol{I}_n & \boldsymbol{0}_{n,p} & \boldsymbol{0}_{n,3r}
\end{pmatrix}
\end{align}
is a $(2n+p+3r)\times (2n+p+3r)$ matrix. Therefore, $\boldsymbol{H}$ and $\boldsymbol{Q}$ satisfy $\boldsymbol{H}'\boldsymbol{Q}=\boldsymbol{0}_{n+p+3r,n}$ and $\boldsymbol{Q}\boldsymbol{Q}'=\boldsymbol{I}_{2n+p+3r}-\boldsymbol{H}(\boldsymbol{H}'\boldsymbol{H})^{-1}\boldsymbol{H}'$. This particular structure will be motivated in the next section, which produces a familiar form for signal-to-noise covariance. For prediction, we filter across fine-scale and discrepancy terms. That is, we use posterior summaries of $\boldsymbol{x}_m(\cdot)'\boldsymbol{\beta}_m+\boldsymbol{g}^*(\cdot)'\boldsymbol{\eta}$, where $\boldsymbol{g}^*(\boldsymbol{s})'=(\boldsymbol{g}_1(\boldsymbol{s})',\boldsymbol{g}_1(\boldsymbol{s})',\boldsymbol{0}_{1,r})$ when $m=1$, $\boldsymbol{g}^*(A)'=(\boldsymbol{g}_2(A)',\boldsymbol{0}_{1,r},\boldsymbol{g}_2(A)')$ when $m=2$, and $\boldsymbol{g}^*(\boldsymbol{s})'=(\boldsymbol{g}_3(\boldsymbol{s})',\boldsymbol{0}_{1,r},\boldsymbol{0}_{1,r})$ when $m=3$.

\subsection{Technical Development}

We combine the observed data in (\ref{eq:datamodel}) into a single vector $\mathbf{z}$ such that $\mathbf{z}=(\mathbf{z}_1',\mathbf{z}_2',\mathbf{z}_3')'$ where $\mathbf{z}_1=(Z_1(\boldsymbol{s}_1),...,Z_1(\boldsymbol{s}_{n_1^*}))'$, $\mathbf{z}_2=(Z_2(A_1),...,Z_2(A_{n_2}))'$, and $\mathbf{z}_3=(Z_3(\boldsymbol{s}_1),...,Z_3(\boldsymbol{s}_{n_1}))'$. For the Gaussian distributed $\mathbf{z}_1$ and $\mathbf{z}_2$, we store their variances in diagonal matrices $\boldsymbol{D}_{\sigma,1}=diag(1/\sigma^2_{1,i}:i=1,...,n_1^*)$ and $\boldsymbol{D}_{\sigma,2}=diag(1/\sigma^2_{2,j}:j=1,...,n_2)$, respectively. Availing of results from \citet{BC23}, we derive the marginal posterior distribution of $\boldsymbol{\xi},\boldsymbol{\beta},\boldsymbol{\eta}$, and $\boldsymbol{q}$. This derivation does not follow immediately from \citet{BC23}. In particular, \citet{BC23} did not jointly model Gaussian and Bernoulli data, and modeled spatial data on a single scale. However, it follows (see Appendix A) that 
\begin{align}\label{eq:GCM:DT}
(\boldsymbol{\xi}',\boldsymbol{\beta}',\boldsymbol{\eta}',\boldsymbol{q}')'|\mathbf{z}\sim GCM(\boldsymbol{\alpha},\boldsymbol{\kappa},\boldsymbol{0}_{2n+p+3r,1},\boldsymbol{V},\pi,\boldsymbol{D};\boldsymbol{\psi}),
\end{align}
 where $\boldsymbol{\alpha}=(\mathbf{z}_1'\boldsymbol{D}_{\sigma,1}',\mathbf{z}_2'\boldsymbol{D}_{\sigma,2}',\mathbf{z}_3'+\alpha_{\xi}\boldsymbol{1,n_1},\boldsymbol{0}_{1,n+p+3r})'$, $\boldsymbol{\kappa}=(\frac{1}{2}\boldsymbol{1}_{1,n_1^*}\boldsymbol{D}_{\sigma,1}',\frac{1}{2}\boldsymbol{1}_{1,n_2}\boldsymbol{D}_{\sigma,2}',\boldsymbol{1}_{1,n_1}+2\alpha_{\xi}\boldsymbol{1}_{1,n_1},\frac{1}{2}\boldsymbol{1}_{1,n+p+3r})'$, $\boldsymbol{0}_{2n+p+3r,1}$ is a $(2n+p+3r)$-dimensional column vector of zeros, $\boldsymbol{V}^{-1}=(\boldsymbol{H},\boldsymbol{Q})$ where $\boldsymbol{H}$ and $\boldsymbol{Q}$ are defined in (\ref{eq:Hmatrix}), $\pi(\boldsymbol{\theta})$ are the proper prior distributions for the hyperparameters $\boldsymbol{\theta}$, $\boldsymbol{D}(\boldsymbol{\theta})^{-1}=blkdiag(\boldsymbol{I}_n,\boldsymbol{D}_{\beta}(\boldsymbol{\theta})^{-1},\boldsymbol{D}_{\eta}(\boldsymbol{\theta})^{-1},\frac{1}{\sigma_\xi}\boldsymbol{I}_n)$, and the unit-log partition function $\boldsymbol{\psi}(\boldsymbol{h})=(\psi_G(h_{1,1}),...,\psi_G(h_{1,n_1^*}),\psi_G(h_{2,1}),...,\psi_G(h_{2,n_2}),\psi_B(h_{3,1}),...,\psi_B(h_{3,n_1}),$\newline$\psi_G(h^*_{1,1}),...,\psi_G(h^*_{1,n_1^*}),\psi_G(h^*_{2,1}),...,\psi_G(h^*_{2,n_2}),\psi_G(h^*_{3,1}),...,\psi_G(h^*_{3,n_1}),\psi_G(h^*_{\beta,1}),...,\psi_G(h^*_{\beta,p}),$\newline$\psi_G(h^*_{\eta,1}),...,\psi_G(h^*_{\eta,3r}))'$ for any $\boldsymbol{h}=(h_{1,1},...,h_{1,n_1^*},h_{2,1},...,h_{2,n_2},h_{3,1},...,h_{3,n_1},h^*_{1,1},...,h^*_{1,n_1^*},$\newline$h^*_{2,1},...,h^*_{2,n_2},h^*_{3,1},...,h^*_{3,n_1},h^*_{\beta,1},...,h^*_{\beta,p},h^*_{\eta,1},...,h^*_{\eta,3r})'\in \mathbb{R}^{2n+p+3r}$.

Equation (\ref{eq:GCM:DT}) is particularly powerful because we know how to sample independent replicates, without MCMC and without approximations of the posterior distribution, from the density function with $\boldsymbol{\zeta}=(\boldsymbol{\xi}',\boldsymbol{\beta}',\boldsymbol{\eta}')'$ via Equation (\ref{eq:transformation}). Specifically, we compute 
\begin{equation}
\begin{aligned}
    \label{eq:compute}
    \begin{pmatrix}
    \boldsymbol{\zeta} \\
    \boldsymbol{q}
    \end{pmatrix}=\boldsymbol{V}\boldsymbol{D}(\boldsymbol{\theta})\boldsymbol{w}=\begin{pmatrix}
        (\boldsymbol{H}'\boldsymbol{H})^{-1}\boldsymbol{H}' \\
        \boldsymbol{Q}'
    \end{pmatrix}\boldsymbol{D}(\boldsymbol{\theta})\boldsymbol{w},
\end{aligned}
\end{equation}
where $\boldsymbol{\theta}$ is drawn from its proper prior and the elements of the $(2n+p+3r)$-dimensional vector $\boldsymbol{w}$ are independently DY where the $i$-th component is DY with shape and rate found in the $i$-th component of $\boldsymbol{\alpha}$ and $\boldsymbol{\kappa}$.

Traditionally, additive terms such as $\boldsymbol{\delta}_y$ are assumed independent of the signal $\mathbf{y}^*=\boldsymbol{X}\boldsymbol{\beta}+\boldsymbol{G}\boldsymbol{\eta}+\boldsymbol{\xi}$. However, in our observational study several different processes influence the data (e.g., humidity, greenhouse gases, etc.), but are not explicitly modeled in $\mathbf{y}^*$ via covariates or through the implied covariances. Consequently, the discrepancy term $\boldsymbol{\delta}_y$ can not reasonably be assumed independent of $\mathbf{y}^*$. Our model explicitly accounts for this, and the implied posterior produces a parametric form for the cross signal-to-noise covariance. In particular, it is straightforward to show that (see Appendix B),
\begin{align}\label{eq:cross:signal:noise}
cov(\mathbf{y}^*,\boldsymbol{\delta}_y|\boldsymbol{z},\boldsymbol{\theta})=\boldsymbol{J}\boldsymbol{H}(\boldsymbol{H}'\boldsymbol{H})^{-1}\boldsymbol{H}'\boldsymbol{D}(\boldsymbol{\theta})cov(\boldsymbol{w}|\boldsymbol{\alpha},\boldsymbol{\kappa})\boldsymbol{D}(\boldsymbol{\theta})'\{\boldsymbol{I}_{2n+p+3r}-\boldsymbol{H}(\boldsymbol{H}'\boldsymbol{H})^{-1}\boldsymbol{H}'\}\boldsymbol{J}',
\end{align}
where $\boldsymbol{\delta}_y=(\boldsymbol{\delta}_1',\boldsymbol{\delta}_2',\boldsymbol{\delta}_3')'$ with $\boldsymbol{\delta}_1=(\delta_1(\boldsymbol{s}_1),...,\delta_1(\boldsymbol{s}_{n_1^*}))'$, $\boldsymbol{\delta}_2=(\delta_2(A_1),...,\delta_2(A_{n_2}))'$, $\boldsymbol{\delta}_3=(\delta_3(\boldsymbol{s}_1),...,\delta_3(\boldsymbol{s}_{n_1}))'$, $\boldsymbol{J}=(\boldsymbol{I}_n,\boldsymbol{0}_{n,p},\boldsymbol{0}_{n,3r},\boldsymbol{0}_{n,n})$, $\boldsymbol{H}$ is as defined in (\ref{eq:Hmatrix}), $\boldsymbol{D}(\boldsymbol{\theta})^{-1}=blkdiag(\boldsymbol{I}_n,$\newline$\boldsymbol{D}_{\beta}(\boldsymbol{\theta})^{-1},\boldsymbol{D}_{\eta}(\boldsymbol{\theta})^{-1},\frac{1}{\sigma_\xi}\boldsymbol{I}_n)$, and $cov(\boldsymbol{w}|\boldsymbol{\alpha},\boldsymbol{\kappa})$ is a diagonal matrix with $(i,i)$-th entry equal to the variance of the $i$-th element of $\boldsymbol{w}$ in Equation (\ref{eq:compute}).

This particular cross-dependence structure is similar to that of the ordinary least squares (OLS) estimator and the OLS residuals in the heteroskedastic setting. This parametric model for the cross-signal-to-noise dependence is developed similar to \citet{bradley23} and \citet{BC23}, but developed for the multi-type and multiscale setting as the matrix $\boldsymbol{H}$ consists of basis functions defined over different scales and $\boldsymbol{w}$ corresponds to different data types. 
	
\section{Simulation}

 We include a simulation study to illustrate the high inferential and computational performance of our model relative to the BHM that does not leverage discrepancy (or feedback) error. This is done to illustrate that the consequence of including this non-standard term has small practical impact on inferences, but provides enormous computational benefits. We consider both the setting where discrepancy is not present (i.e., $\delta_m=0$) and the setting where discrepancy is present (i.e., $\delta_m\neq 0$). In particular, for both settings, we fit our model with discrepancy term in (\ref{eq:datamodel}) based on Equation (\ref{eq:GCM:DT}), and fit the model in (\ref{eq:datamodel}) assuming $\delta_m=0$ via MCMC. 
 
\begin{figure}
	\centering
	\begin{subfigure}[b]{\textwidth}
		\centering
		\includegraphics[width=0.25\linewidth]{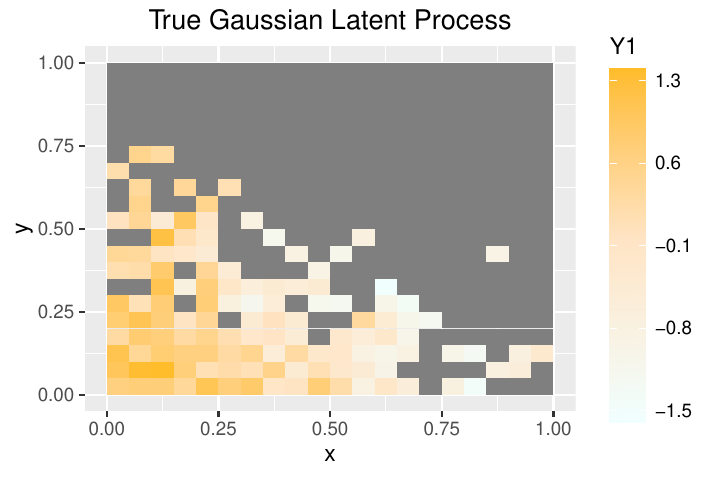}
        \includegraphics[width=0.25\linewidth]{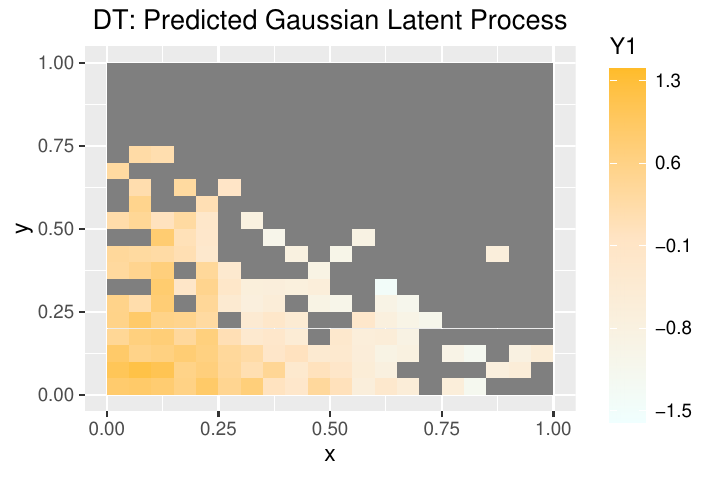} 
        \includegraphics[width=0.25\linewidth]{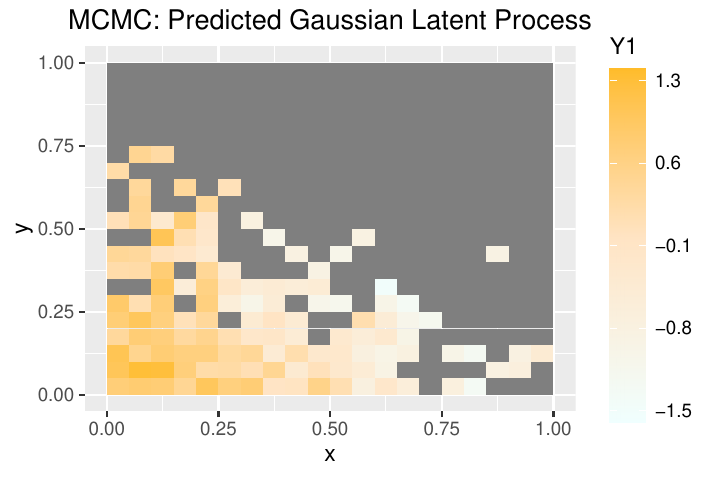}\\
        \includegraphics[width=0.25\linewidth]{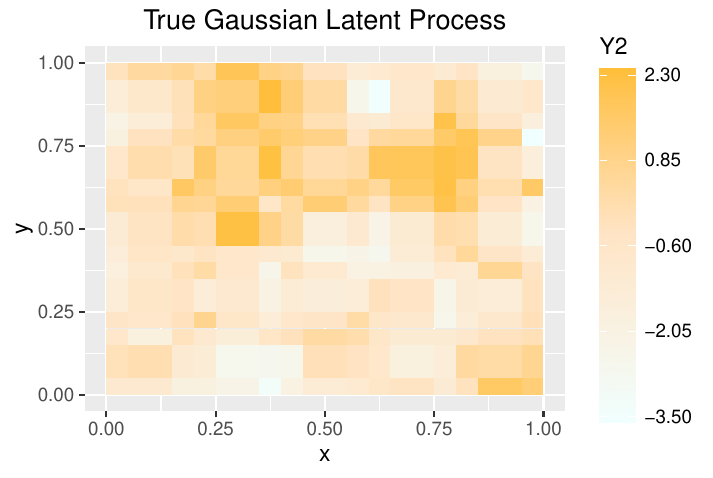} 
        \includegraphics[width=0.25\linewidth]{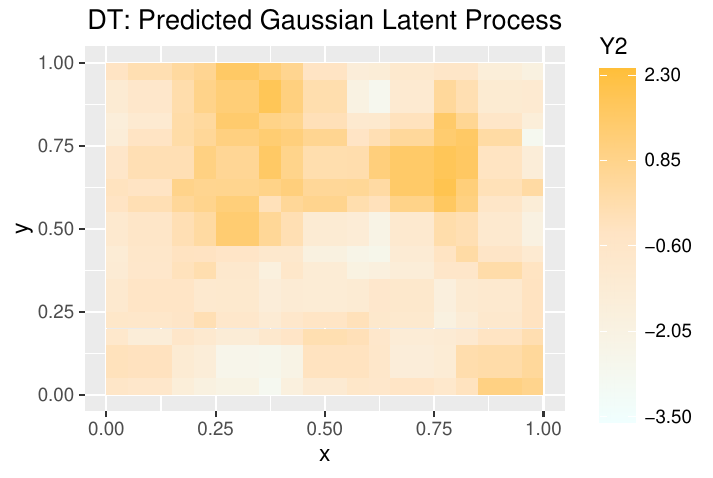} 
        \includegraphics[width=0.25\linewidth]{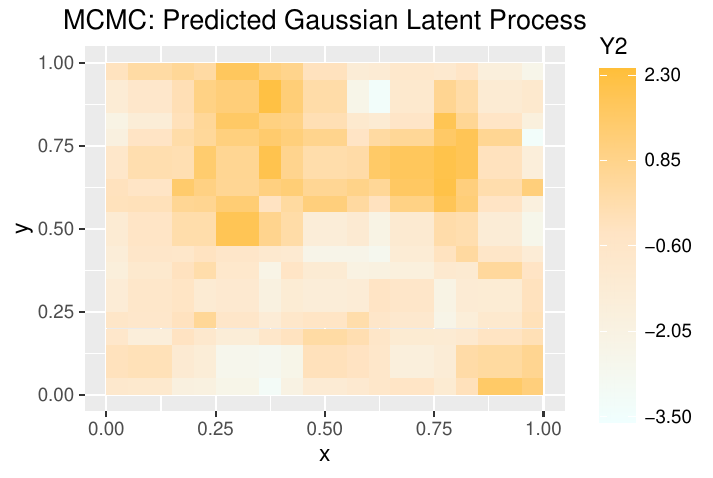}\\
        \includegraphics[width=0.25\linewidth]{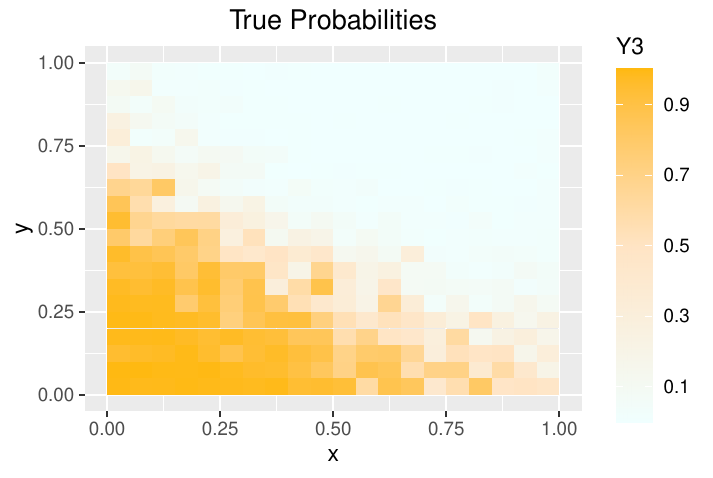} 
        \includegraphics[width=0.25\linewidth]{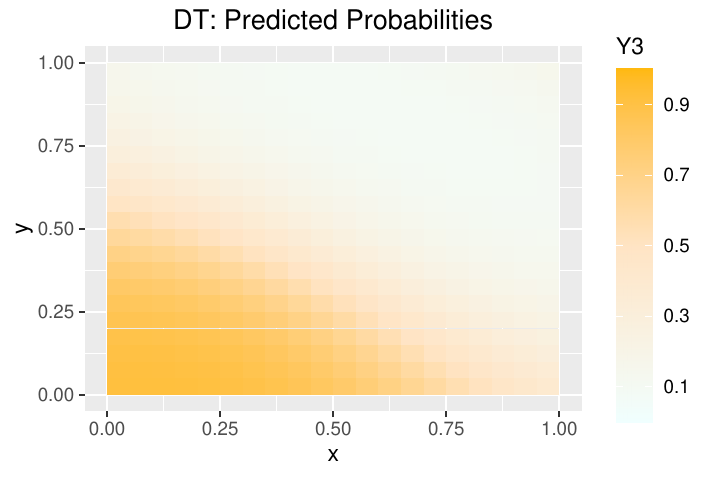} 
        \includegraphics[width=0.25\linewidth]{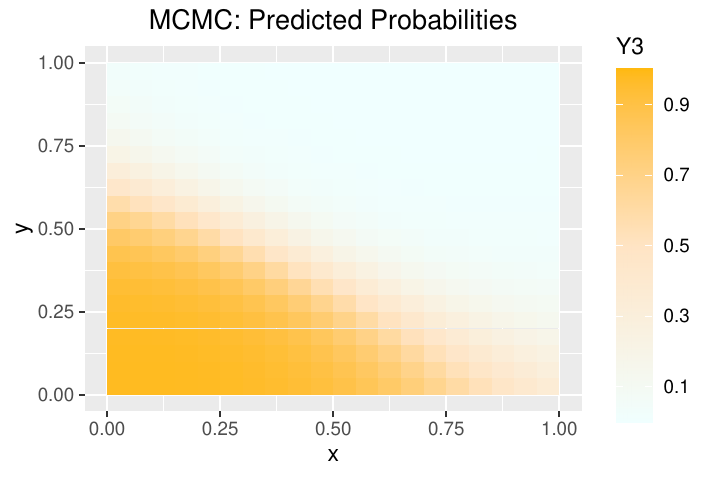}
        \caption{Discrepancy is present}
        \label{fig:simpanel_epr}
	\end{subfigure}

        \vspace{0.1cm}

\begin{subfigure}[b]{\textwidth}
		\centering
		\includegraphics[width=0.25\linewidth]{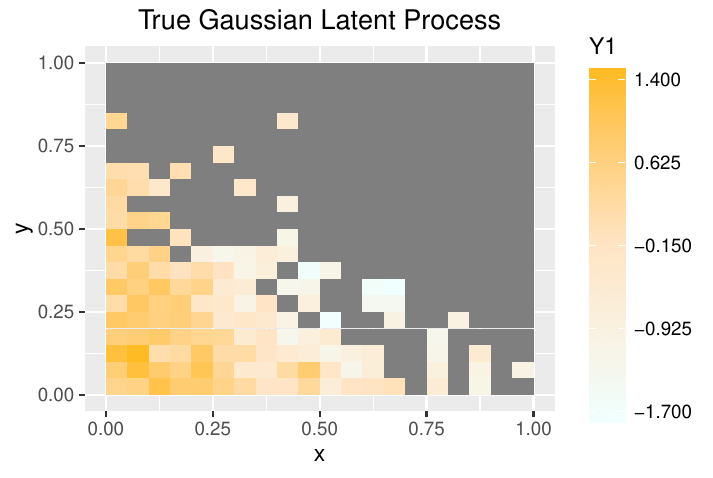}
        \includegraphics[width=0.25\linewidth]{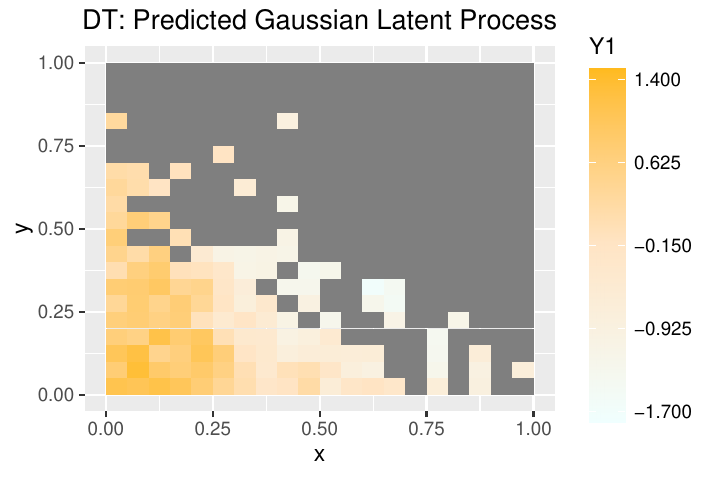} 
        \includegraphics[width=0.25\linewidth]{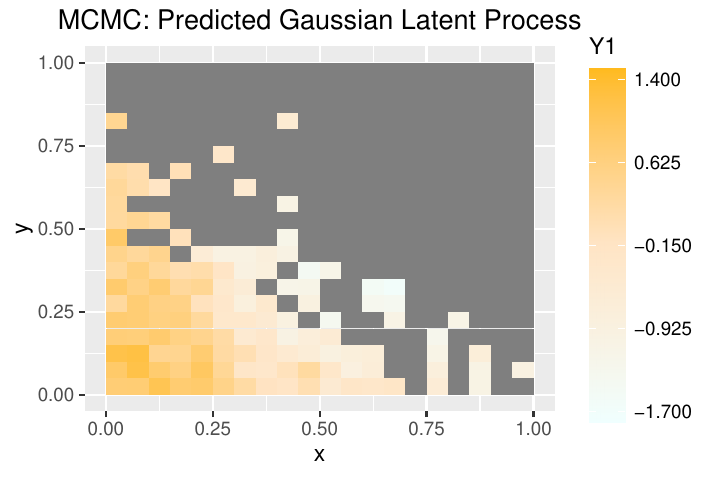}\\
        \includegraphics[width=0.25\linewidth]{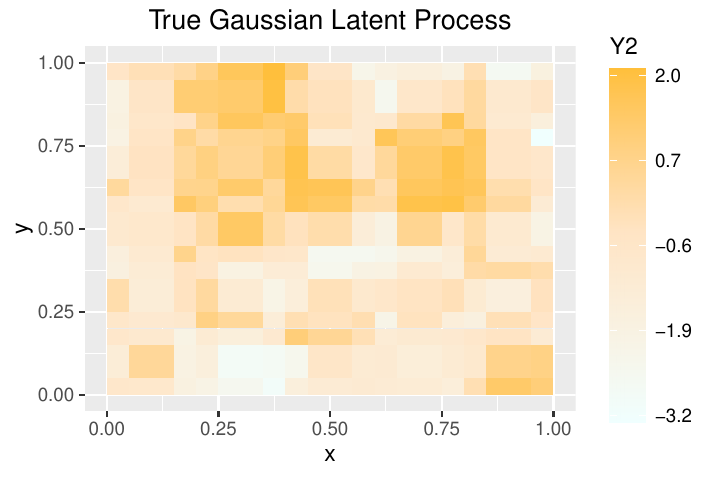} 
        \includegraphics[width=0.25\linewidth]{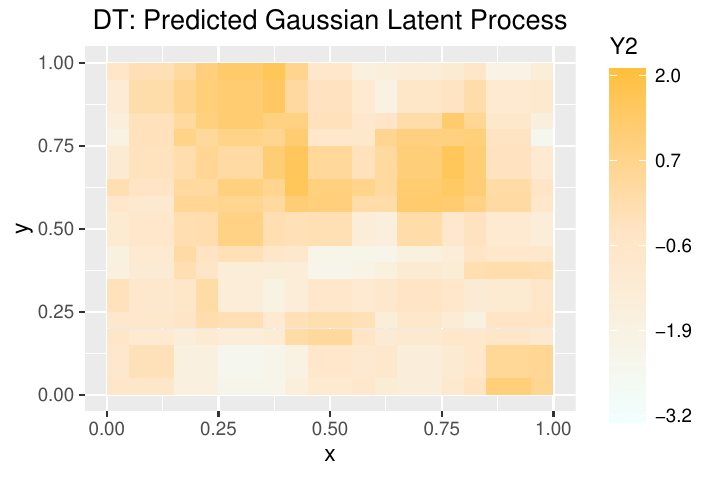} 
        \includegraphics[width=0.25\linewidth]{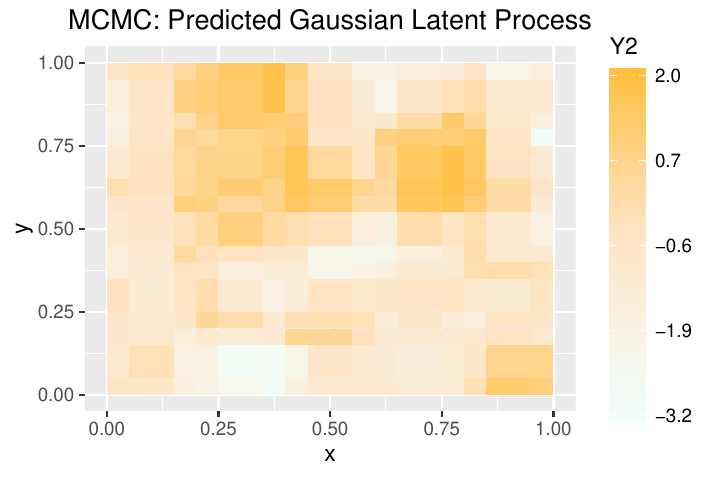}\\
        \includegraphics[width=0.25\linewidth]{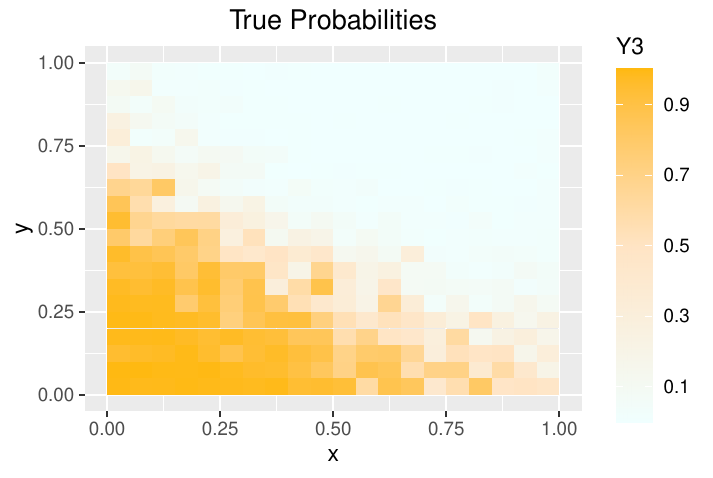} 
        \includegraphics[width=0.25\linewidth]{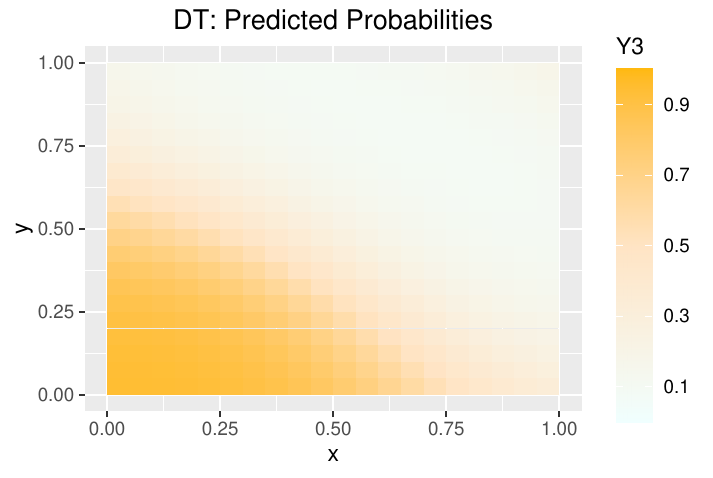} 
        \includegraphics[width=0.25\linewidth]{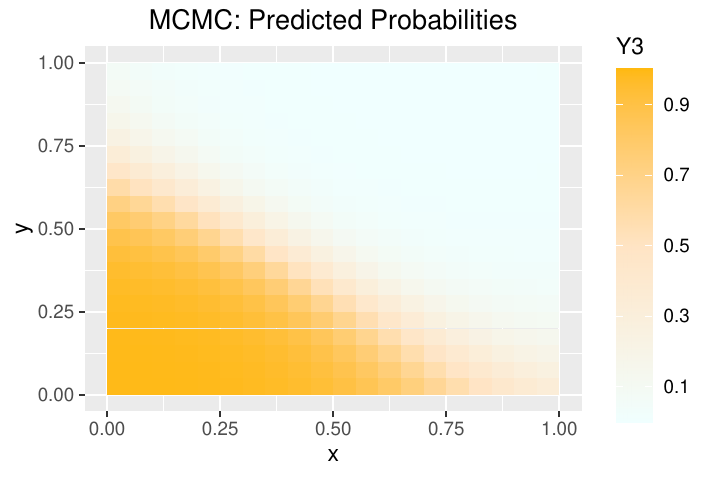}
        \caption{Discrepancy is not present}
        \label{fig:simpanel_mcmc}
	\end{subfigure}
        
	\caption{Comparison between BHM with discrepancy term (DT) and traditional BHM using MCMC when (a) discrepancy is present and (b) discrepancy is not present. The first column presents the simulated latent process for each variable. The second and third columns provide the corresponding prediction obtained from our BHM with discrepancy term and traditional BHM, respectively.}
	\label{fig:simpanel}
\end{figure}

 We generate data on a $20\times 20$ equally spaced point-referenced locations over the unit square (i.e., $[0,1]\times[0,1]$). Areal data are defined on 225 irregular regions over the spatial domain whose union forms the unit square as done in \citet{ZB24}, which is seen in Figure~\ref{fig:simpanel}. The covariates are extracted from the real data by placing a $20\times 20$ square within the boundary of California so that $\boldsymbol{x}(\boldsymbol{s}_i)$ corresponds directly, one-to-one, to the covariates from the real data. The true value of $\beta_m$ is set equal to the posterior means of fixed effects obtained by fitting the BHM with discrepancy term to the real data. The random effects (i.e., $\boldsymbol{\eta}_m$) are assumed Gaussian with mean zero and covariance $\boldsymbol{I}_r$. We use Gaussian radial basis functions and choose $r=50$ equally spaced knots using the \textit{cover.design} function in the R package \textit{fields}. The fine-scale term is assumed Gaussian with mean zero and the variance is chosen so that the signal-to-noise ratio is large, and equal to 20. Similarly, for a given $i$ and $j$, the parameters $\sigma_{1i}^{2}$ and $\sigma_{2j}^{2}$ are chosen to produce signal-to-noise ratios $SNR_{i}$ and $SNR_{j}$. The values of $SNR_{i}$ and $SNR_{j}$ are uniformly drawn from 1 to 10. These specifications define the model that generates the data without a discrepancy parameter.

 To generate data with a discrepancy term, we simulate from the posterior distribution from our discrepancy model using $\mathbf{y}^{*} = \textbf{X}\boldsymbol{\beta} + \textbf{G}\boldsymbol{\eta} + \boldsymbol{\xi}$ as Gaussian data, where $\textbf{X}$, $\boldsymbol{\beta}$, $\textbf{G}$, $\boldsymbol{\eta}$, and $\boldsymbol{\xi}$ are defined above. This produces estimates $\mathbf{y}_{DT} = \textbf{X}\widehat{\boldsymbol{\beta}}_{DT} + \textbf{G}\widehat{\boldsymbol{\eta}}_{DT} + \widehat{\boldsymbol{\xi}}_{DT}$, where $\widehat{\boldsymbol{\beta}}_{DT}$, $\widehat{\boldsymbol{\eta}}_{DT}$, and $\widehat{\boldsymbol{\xi}}_{DT}$ are posterior means from the discrepancy term (DT) model that uses $\mathbf{y}^{*}$ as Gaussian distributed data. Then we set $\boldsymbol{\delta}_{y} = \mathbf{y}_{DT} - \mathbf{y}^{*}$. Recall the discrepancy term is meant to model signal-to-noise covariance. This simulation design, produces such a covariance, since $cov(\mathbf{y}^{*},\boldsymbol{\delta}_{y}) = cov(\mathbf{y}^{*},\mathbf{y}_{DT})-var(\mathbf{y}^{*})$, which is not equal to zero.

We plot the results in Figure \ref{fig:simpanel} for one simulated dataset. Figure \ref{fig:simpanel_epr} shows the results when discrepancy is present in the simulated data, while Figure \ref{fig:simpanel_mcmc} shows the results when discrepancy is not present in the simulated data. In both figures, the first column shows the simulated latent process for $Y_1$, $Y_2$, and $Y_3$, respectively. The second column shows the corresponding prediction obtained by fitting the BHM with discrepancy term, and the third column shows the prediction obtained by fitting the traditional BHM using MCMC. For the first variable with excessive zeros (note that zero-valued locations are in grey), both methods appear to capture the main spatial patterns. For the second variable, both methods perform similarly well. For the third (Bernoulli) variable, the discrepancy model fit seems to have slightly smoother predicted probabilities than MCMC while still showing similar spatial patterns. These conclusions appear to remain valid regardless of whether the discrepancy is present based on this one particular illustration.

To more rigorously compare the two methods, we simulate 100 replicates of data in both settings and compute the average of mean square prediction error (MSPE) between the simulated latent process and predicted latent process and the continuous rank probability score (CRPS, \citealp{GK14}) for the first and second variables. Recall for the discrepancy model $\boldsymbol{\delta}_{y}$ is interpreted as a noise parameter, and as such, the true latent process is $\boldsymbol{y}_{DT}$, whereas for the model without a discrepancy parameter the true latent process is $\boldsymbol{y}^*$. For the Bernoulli variable $Y_3$, we compute the Hellinger's distance \citep{hellinger} between the true simulated probabilities and the predicted probabilities. Additionally, we include the average of mean squared error (MSE) between the true/realized and estimated fixed/random effects and central processing unit (CPU) time in seconds. In terms of predictive accuracy, our fitted discrepancy BHM appears to perform better when discrepancy is present (see Table \ref{tab:eprtab}), while the traditional BHM fitted via MCMC appears to perform better when discrepancy is not present (see Table \ref{tab:mcmctab}). In general, both methods exhibit comparable inferential performance, with one method slightly outperforming the other depending on the dataset, particularly whether discrepancy is present. However, our method is significantly more efficient computationally, taking less than 50 seconds on average, which is approximately 26 times faster than the traditional MCMC method.

\begin{table}[!ht]
    \begin{subtable}[h]{1.0\textwidth}
	\centering
	\begin{tabular}{c@{\hskip 0.2in} c@{\hskip 0.2in} c@{\hskip 0.2in} c}
		\hline
		\hline
		&  & DT & MCMC \\
		\hline
		\multirow{2}{*}{$Y_1$} & MSPE & 1.090 (0.068) & 1.111 (0.057) \\
		 & CRPS & 0.646 (0.118) & 0.744 (0.027) \\
		 \hline
		 \multirow{2}{*}{$Y_2$} & MSPE & 1.184 (0.056) & 1.209 (0.069) \\
		 & CRPS & 0.713 (0.216) & 0.756 (0.034) \\
		\hline 
		$Y_3$ & HD &  3.493 (0.113) & 4.597 (0.280) \\
		\hline
		& CPU & 49.435 (1.603) & 1282.710 (46.038) \\
		& MSE & 0.917 (0.021) & 0.979 (0.029) \\
		\hline
		\hline
	\end{tabular}
        \caption{Discrepancy is present}
        \label{tab:eprtab}
 \end{subtable}

 \vspace{1cm}
 \begin{subtable}[h]{1.0\textwidth}
	\centering
	\begin{tabular}{c@{\hskip 0.2in} c@{\hskip 0.2in} c@{\hskip 0.2in} c}
		\hline
		\hline
		&  & DT & MCMC \\
		\hline
		\multirow{2}{*}{$Y_1$} & MSPE & 0.083 (0.031) & 0.054 (0.016) \\
		 & CRPS & 0.377 (0.279) & 0.128 (0.019) \\
		 \hline
		 \multirow{2}{*}{$Y_2$} & MSPE & 0.148 (0.040) & 0.152 (0.045) \\
		 & CRPS & 0.513 (0.381) & 0.218 (0.037) \\
		\hline 
		$Y_3$ & HD & 1.858 (0.084)  & 1.198 (0.066) \\
		\hline
		& CPU & 47.444 (1.549) & 1240.106 (34.503)\\
		& MSE & 0.737 (0.169) & 0.558 (0.025) \\
		\hline
		\hline
	\end{tabular}
        \caption{Discrepancy is not present}
        \label{tab:mcmctab}
 \end{subtable}
	\caption{\label{tab:eprvsmcmc}Simulation study results comparing BHM with discrepancy term (DT) and traditional BHM fitted via MCMC, when (a) discrepancy is present and (b) discrepancy is not present. For the Gaussian-distributed $Y_1$ and $Y_2$, we report the mean squared prediction error (MSPE) and continuous rank probability score (CRPS). For the Bernoulli-distribued $Y_3$, we report the Hellinger's distance between the simulated probabilities and predicted probabilities. The mean squared error (MSE) between the true and estimated fixed and random effects and central processing unit (CPU) time in seconds are also included. The values shown are the means (standard deviations) over the 100 simulated replicates of dataset.}
\end{table}

	\section{Application}

 Within our spatial domain of interest (i.e., contiguous United States), there are 80,817 observed point-referenced locations excluding those fall in the islands and water areas for the fire and fire indicator variables, out of which 65,538 locations have no fire (i.e., $Y_1=0$ and $Y_3=0$ at these locations). For the population change data, all 3,109 counties (including county-equivalents) of the contiguous U.S. are observed with no missing data. We plot the annual average of number of fires (see Figure~\ref{fig:fire_data}) and annual population change in percentage (see Figure~\ref{fig:pop_data}) in the United States from July 2020 to June 2021. From Figure~\ref{fig:fire_data}, we see that fires predominantly occur along the west coast, central, and southern regions of the U.S. Based on the color scale (recall that the values shown are log-transformed for better visualization), northern California exhibits the largest annual average of number of fires. Spatial clustering patterns are also detected in the central area around Kansas and Missouri, and in the southeast around Georgia, Alabama, and Florida. Large grey areas are seen in the Southwest as well as the Northeast of the U.S., suggesting these regions experience few to no fires. In Figure~\ref{fig:pop_data}, the population change data present a subtler spatial pattern compared to the fire data. The map indicates a mixture of population growth and decline across the counties. From these data plots alone, it is hard to tell whether there is a direct correlation between the two responses, and hence, further analysis is required to jointly study these multiscale responses.

 We apply our BHM with a discrepancy term to the fire and population change dataset. We obtain four point-referenced covariates from NASA for the fire variable \citep{nasa21}, namely, land surface temperature, rainfall, vegetation index, and elevation. We also use county-referenced median household income data from the ACS as a covariate for the population change variable \citep{acs24}. To investigate the effects of these covariates on their corresponding responses, we present the posterior means and 95\% credible intervals (CI) for the coefficients in Table~\ref{tab:coeff}. Among the four covariates for fire, land surface temperature appears to have a significant (i.e., the 95\% CI excludes zero) negative effect on the annual average of number of fires. This finding might seem counterintuitive, as higher land surface temperatures can dry out vegetation, making it more flammable and likely to ignite. However, unmeasured confounders may explain this. The remaining three covariates for fire are found insignificant as their 95\% CIs contain zero. For population change, the median household income has an associated coefficient that is significantly positive, suggesting that the population in a county is expected to increase as the median household income of that county increases.

 \begin{table}
    \begin{subtable}[h]{1.0\textwidth}
    \centering
    \begin{tabular}{cccc}
    \hline
    \hline
      Response   & Covariates & Posterior Mean  & 95\% CI \\
      \hline
        \multirow{5}{*}{Fire} & Intercept & 1.170 & (-0.969,3.485)\\
         & Land Surface Temperature ($\times 10^{-2}$) & -4.854 & (-7.116,-1.391)\\
         & Rainfall ($\times 10^{-3}$) & 0.786 & (-5.520,9.169) \\
         & Vegetation Index & -0.666 & (-1.355,0.040) \\
         & Elevation ($\times 10^{-5}$) & 0.478 & (-1.049,2.136) \\
        \hline
        \multirow{3}{*}{Population Change} & Intercept & -1.539 & (-3.446,0.276) \\
         & Median Household Income ($\times 10^{-5}$) & 2.416 & (2.021,3.052)\\
         \hline
        Fire Indicator & Intercept & -1.786 & (-2.520,-1.088)\\
        \hline
        \hline
    \end{tabular}
    \caption{Posterior mean and 95\% credible interval (CI) of coefficients}
    \label{tab:coeff}
    \end{subtable}
    \vspace{1cm}

    \begin{subtable}[h]{1.0\textwidth}
    \centering
    \begin{tabular}{cccc}
    \hline
    \hline
        & Fire & Population Change  & Fire Indicator \\
      \hline
        MSPE &  0.089 & 0.559   & ---  \\
        IS & 1.875  &  7.818  & ---  \\
        AUC & ---  &  ---  &  0.783 \\
        \hline
        \hline
    \end{tabular}
    \caption{MSPE, IS, and AUC by responses}
    \label{tab:predacry}
    \end{subtable}

    \caption{\label{tab:realresult}Real data analysis results. In (a), we show the posterior mean and 95\% credible interval for each coefficient. In (b), we show the mean squared prediction error (MSPE) and interval score (IS) for the fire and population change variables. For the fire indicator variable, we present the area under the curve (AUC) of the ROC curve.}
\end{table}

 To evaluate the predictive accuracy of our model, we report the MSPE and interval score (IS) for the fire and population change variables in Table \ref{tab:predacry}. The MSPE in the real analysis is computed between the observed data and the predicted latent process. The IS assesses the accuracy and uncertainty of the predictions by penalizing both the width of the $(1-\alpha)\times100\%$ prediction interval (in our case, $\alpha=0.05$) and the failure to cover the observed value \citep{GR07}. Results show that the point-referenced fire variable has a smaller MSPE and IS, which is expected given its finer scale information and larger number of observations compared to population change. For the Bernoulli-distributed fire indicator variable, we report the area under the curve (AUC) of the Receiver Operating Characteristic (ROC) curve \citep{Mandrekar10}. The value of the AUC ranges from 0 to 1, where an AUC value of 1 indicates perfect classification, 0.5 suggests performance equivalent to random guessing, and less than 0.5 indicates performance worse than random guessing. With an AUC of 0.783, our model shows a good discriminative ability to differentiate between occurrences of fire and no fire. Please see Appendix C for the ROC curve.

\begin{figure}[!ht]
		\centering
		\includegraphics[width=\linewidth]{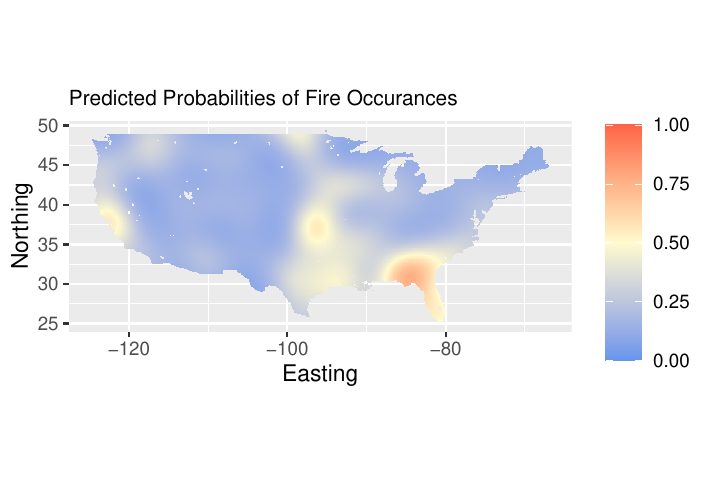} 
	\caption{The predicted probabilities of fire occurrences over 80,817 point-referenced locations in the contiguous U.S.}
	\label{fig:fireprob_eprpred}
\end{figure}

 The inclusion of the Bernoulli-distributed fire indicator variable allows us to predict the probabilities of fire occurrences for the U.S., which is important since effective fire management and prevention strategies heavily rely on accurate risk assessment. Understanding the probability of fire occurrences helps allocate resources efficiently, informs policy decisions, and enhances public safety by enabling timely interventions high-risk areas. We give the plot of predicted probabilities of fire occurrences in Figure~\ref{fig:fireprob_eprpred}. This plot is quite smooth, however, upon inspection of Figure~\ref{fig:fire_data}, we see that these predicted values mimic the smooth patterns in the data. Notably, while the central and west coast regions show moderate-to-high probabilities of fire, a large area in the southeast seems to suffer from the highest risk of burning. The prediction plots for the annual average number of fires and population change are also available (see Appendix D for details).

\begin{figure}
	\centering
	\begin{subfigure}[t]{0.5\textwidth}
		\centering
		\includegraphics[width=\linewidth]{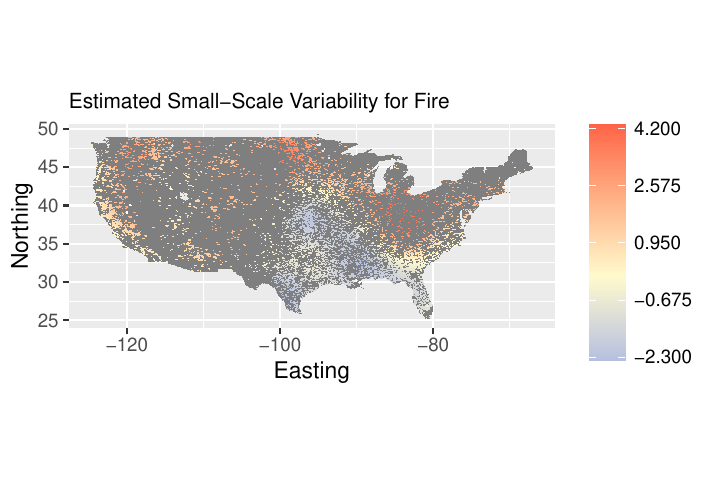} 
		\caption{}
		\label{fig:fire_basis}
	\end{subfigure}
	\begin{subfigure}[t]{0.5\textwidth}
		\centering
		\includegraphics[width=\linewidth]{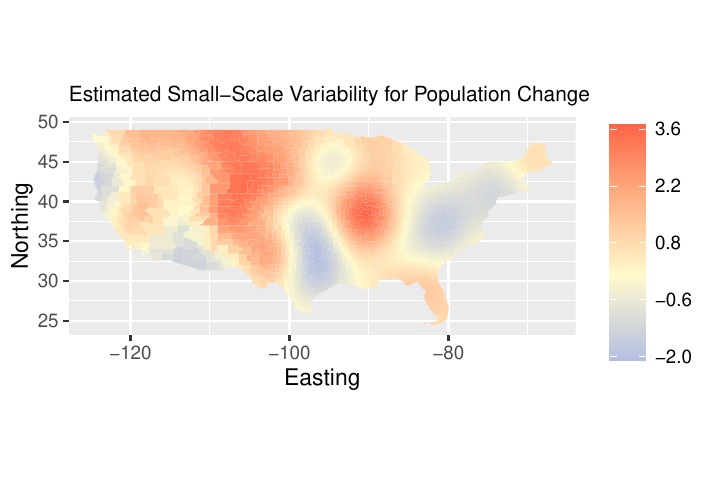} 
		\caption{}
		\label{fig:pop_basis}
	\end{subfigure}

    \vspace{0.1cm}
    \begin{subfigure}[t]{0.5\textwidth}
		\centering
		\includegraphics[width=\linewidth]{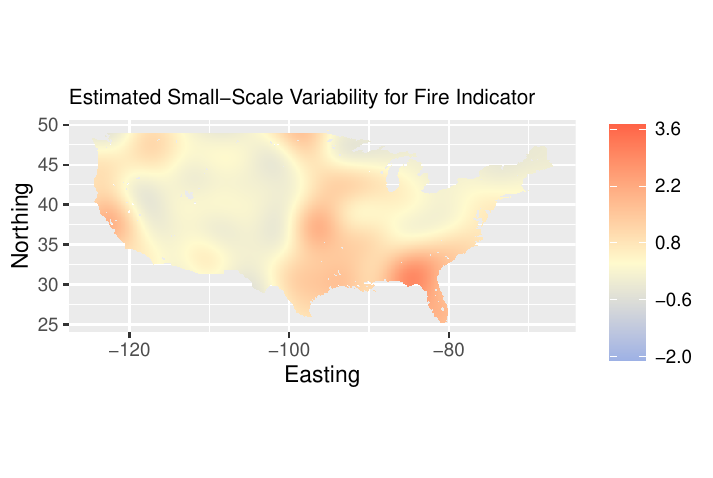} 
		\caption{}
		\label{fig:fireprob_basis}
	\end{subfigure}
	\caption{Plots of estimated small-scale variability terms with shared random vector (i.e., $\boldsymbol{g}_1(\boldsymbol{s}_i)'\hat{\boldsymbol{\eta}}_1$, $\boldsymbol{g}_2(A_j)'\hat{\boldsymbol{\eta}}_1$, $\boldsymbol{g}_3(\boldsymbol{s}_k)'\hat{\boldsymbol{\eta}}_1$ where $\hat{\boldsymbol{\eta}}_1$ is the posterior mean of $\boldsymbol{\eta}_1$) for (a)fire, (b)population change, and (c)fire indicator variables.}
	\label{fig:firenpop_basis}
\end{figure}

 Recall that in our model the correlation between the responses in induced through a random vector $\boldsymbol{\eta}_1$ that is shared across all variables. To investigate this potential correlation, we plot the estimated small-scale variability terms $\boldsymbol{g}_1(\boldsymbol{s}_i)'\hat{\boldsymbol{\eta}}_1$, $\boldsymbol{g}_2(A_j)'\hat{\boldsymbol{\eta}}_1$, and $\boldsymbol{g}_3(\boldsymbol{s}_k)'\hat{\boldsymbol{\eta}}_1$, where $\hat{\boldsymbol{\eta}}_1$ is the posterior mean of $\boldsymbol{\eta}_1$, in Figure \ref{fig:firenpop_basis}. We find that Figures \ref{fig:fire_basis} and \ref{fig:fireprob_basis} appear to have consistent spatial patterns, which is expected given that $\boldsymbol{g}_3(\boldsymbol{s}_k)'\hat{\boldsymbol{\eta}}_1$ for the fire indicator variable should have a strong positive effect in regions where fires are observed (i.e., areas that are not grey in Figure \ref{fig:fire_basis}). In contrast, Figures \ref{fig:pop_basis} and \ref{fig:fireprob_basis} show converse spatial patterns in the Midwest to Western regions of the U.S., suggesting a negative correlation between fire probability and population change. Specifically, in these regions, lower/higher probabilities of fire correspond to greater population increases/decreases. However, this negative correlation pattern is not observed in the Eastern U.S., indicating a potential regional variation in the factors influencing both responses.

 In an effort to fit the traditional BHM using MCMC on our dataset, we encountered significant computational challenges. The implementation with \textit{RJAGS}, which was used successfully in our simulations, became infeasible as R consistently froze and failed to initiate the process. To address this, we developed custom code to increase flexibility and optimize computation. Specifically, we implemented a Metropolis-within-Gibbs algorithm and used an adaptive approach to tune the proposal densities for parameters without closed-form full-conditional distributions. Despite these efforts, the traditional MCMC implementation took approximately 16 hours to generate 2 chains with 10,000 samples each, yet convergence was not achieved after 10,000 iterations (see Gelman-Rubin statistics in Appendix E). In contrast, our discrepancy model completed the analysis in approximately 8 hours.

	\section{Discussion}
	
	In this paper, we propose a novel multivariate multiscale BHM that incorporates a discrepancy term to account for cross-signal-to-noise dependence, while also providing significant computational advantages. We present theoretical work demonstrating that we can bypass MCMC by directly sampling independent replicates from the exact posterior distribution, which dramatically improves the computational speed and is crucial in high-dimensional settings. Moreover, the discrepancy term leads to signal-to-noise dependence, which is expected in wildfire application. Motivated by wildfire and population change data, our model enables efficient joint modeling of point-reference and areal spatial data and accounts for zero-inflation by incorporating a Bernoulli-distributed fire indicator variable, allowing for multi-type modeling.

    We conducted a comprehensive simulation study, considering scenarios both with and without the presence of discrepancy. In both cases, we fit our BHM with discrepancy term, and compare it to a traditional MCMC approach without the discrepancy term. We evaluated both methods in terms of predictive accuracy and computational efficiency. Results show that both models perform comparably in predictive accuracy. However, our discrepancy BHM model has a substantial computational advantage while maintaining similar inferential performance. In our application, we identified several key covariates, such as land surface temperature for fire and median household income for population change. We also produced fire probability predictions across the contiguous U.S., finding that the Southeast has the highest probability of fire occurrences, with elevated risks in the central and western regions. Additionally, we observed a negative correlation between fire probability and population change in the Midwest to western U.S.

    Our discrepancy BHM can easily be adapted to other multivariate multiscale spatial settings. While this paper focuses on Gaussian and Bernoulli-distributed data, the model can be extended to accommodate other distributions. For instance, \citet{BC23} explored Poisson and Binomial settings in univariate, single-scale contexts, which can be generalized to multivariate, multiscale frameworks. Furthermore, while we focus on joint modeling of point-referenced and areal data, our approach can be extended to other multiscale scenarios, such as analyzing multiscale spatial data observed from multiple misaligned areal supports or performing inference at scales different from the data's observation scale.

	\section*{Acknowledgments}
	
	This research was partially supported by the U.S. National Science Foundation (NSF) under NSF grant DMS-2310756.

\clearpage

    \begin{appendices}

    \section{Proof of Equation (\ref{eq:GCM:DT})}

  In this appendix, we present the proof of Equation (\ref{eq:GCM:DT}). We want to show $f(\boldsymbol{\xi},\boldsymbol{\beta},\boldsymbol{\eta},\boldsymbol{q}|\mathbf{z})$ is the GCM stated in Theorem 1. Using Bayes' Rule, we have $f(\boldsymbol{\xi},\boldsymbol{\beta},\boldsymbol{\eta},\boldsymbol{\theta},\boldsymbol{q}|\mathbf{z})\propto$
  \begin{align}
f(\mathbf{z}|\boldsymbol{\xi},\boldsymbol{\beta},\boldsymbol{\eta},\boldsymbol{\theta},\boldsymbol{q})f(\boldsymbol{\xi}|\boldsymbol{\beta},\boldsymbol{\eta},\boldsymbol{\theta},\boldsymbol{q})f(\boldsymbol{\beta}|\boldsymbol{\theta},\boldsymbol{q})f(\boldsymbol{\eta}|\boldsymbol{\theta},\boldsymbol{q})f(\boldsymbol{q})\pi(\boldsymbol{\theta}).
    \label{eq:product}
  \end{align}
  Given the data models in (6) from the main text, we can write the first term in (\ref{eq:product}) as
\begin{align}
f(\mathbf{z}|\boldsymbol{\xi},\boldsymbol{\beta},\boldsymbol{\eta},\boldsymbol{\theta},\boldsymbol{q})=N\exp\left[\boldsymbol{a}_D'\left\{(\boldsymbol{I}_n,\boldsymbol{X},\boldsymbol{G})\begin{pmatrix}
\boldsymbol{\xi} \\
\boldsymbol{\beta} \\
\boldsymbol{\eta} 
\end{pmatrix}-\boldsymbol{\delta}_y\right\}-\boldsymbol{b}_D'\boldsymbol{\psi}_D\left\{(\boldsymbol{I}_n,\boldsymbol{X},\boldsymbol{G})\begin{pmatrix}
\boldsymbol{\xi} \\
\boldsymbol{\beta} \\
\boldsymbol{\eta} 
\end{pmatrix}-\boldsymbol{\delta}_y\right\}\right],
\label{eq:term1}
\end{align}
where $N=\left(\frac{1}{2\pi}\right)^{(n_1^*+n_2)/2}\prod_{i=1}^{n_1^*}\frac{\exp(-Z_1(\boldsymbol{s}_i)^2/2\sigma_{1,i}^2)}{\sigma_{1,i}}\prod_{j=1}^{n_2}\frac{\exp(-Z_2(A_j)^2/2\sigma_{2,j}^2)}{\sigma_{2,j}}$, $\boldsymbol{a}_D=(\mathbf{z}_1'\boldsymbol{D}_{\sigma,1}',\mathbf{z}_2'\boldsymbol{D}_{\sigma,2}',$\newline$\mathbf{z}_3')'$, $\boldsymbol{b}_D=(\frac{1}{2}\boldsymbol{1}_{1,n_1^*}\boldsymbol{D}_{\sigma,1}',\frac{1}{2}\boldsymbol{1}_{1,n_2}\boldsymbol{D}_{\sigma,2}',\boldsymbol{1}_{1,n_1})'$, and $\boldsymbol{\psi}_D(\boldsymbol{h}_D)=(\psi_G(h_{1,1},...,\psi_G(h_{1,n_1^*}),\psi_G(h_{2,1}),...,$\newline$\psi_G(h_{2,n_2}),\psi_B(h_{3,1}),...,\psi_B(h_{3,n_1}))'$ for any $\boldsymbol{h}_D=(h_{1,1},...,h_{1,n_1^*},h_{2,1},...,h_{2,n_2},h_{3,1},...,h_{3,n_1})\in \mathbb{R}^n$. Now, the model for the fine-scale variability term in (\ref{eq:product}) can be written as 
\begin{equation}
\begin{aligned}
f(\boldsymbol{\xi}|\boldsymbol{\beta},\boldsymbol{\eta},\boldsymbol{\theta},\boldsymbol{q})=N_{\xi}\exp\left[\boldsymbol{\alpha}_\xi'\left\{\begin{pmatrix}
\boldsymbol{I}_n & \boldsymbol{X} & \boldsymbol{G} \\
\frac{1}{\sigma_{\xi}}\boldsymbol{I}_n & \boldsymbol{0}_{n,p} & \boldsymbol{0}_{n,3r}
\end{pmatrix}\begin{pmatrix}
\boldsymbol{\xi} \\
\boldsymbol{\beta} \\
\boldsymbol{\eta} 
\end{pmatrix}-\begin{pmatrix}
\boldsymbol{\delta}_y \\
\boldsymbol{\delta}_{\xi}
\end{pmatrix}\right\}\right. \\
\left.-\boldsymbol{\kappa}_{\xi}'\boldsymbol{\psi}_{\xi}\left\{\begin{pmatrix}
\boldsymbol{I}_n & \boldsymbol{X} & \boldsymbol{G} \\
\frac{1}{\sigma_{\xi}}\boldsymbol{I}_n & \boldsymbol{0}_{n,p} & \boldsymbol{0}_{n,3r}
\end{pmatrix}\begin{pmatrix}
\boldsymbol{\xi} \\
\boldsymbol{\beta} \\
\boldsymbol{\eta} 
\end{pmatrix}-\begin{pmatrix}
\boldsymbol{\delta}_y \\
\boldsymbol{\delta}_{\xi}
\end{pmatrix}\right\}\right],
\label{eq:term2}
\end{aligned}
\end{equation}
where $N_{\xi}=\left(\frac{1}{2\pi\sigma^2_{\xi}}^{n/2}\right)$, $\boldsymbol{\alpha}_{\xi}=(\boldsymbol{0}_{1,n_1^*},\boldsymbol{0}_{1,n_2},\alpha_{\xi}\boldsymbol{1}_{1,n_1},\boldsymbol{0}_{1,n})'$, $\boldsymbol{\kappa}_{\xi}=(\boldsymbol{0}_{1,n_1^*},\boldsymbol{0}_{1,n_2},2\alpha_{\xi}\boldsymbol{1}_{1,n_1},\frac{1}{2}\boldsymbol{1}_{1,n})'$, and $\boldsymbol{\psi}_{\xi}(\boldsymbol{h}_{\xi})=(\psi_G(h_{1,1}),...,\psi_G(h_{1,n_1^*}),\psi_G(h_{2,1}),...,\psi_G(h_{2,n_2}),\psi_B(h_{3,1}),...,\psi_B(h_{3,n_1}),\psi_G(h^*_{1,1}),$\newline$...,\psi_G(h^*_{1,n_1^*}),\psi_G(h^*_{2,1}),...,\psi_G(h^*_{2,n_2}),\psi_G(h^*_{3,1}),...,\psi_G(h^*_{3,n_1}))'$ for any $\boldsymbol{h}_{\xi}=(h_{1,1},...,h_{1,n_1^*},h_{2,1},$\newline$...,h_{2,n_2},h_{3,1},...,h_{3,n_1},h^*_{1,1},...,h^*_{1,n_1^*},h^*_{2,1},...,h^*_{2,n_2},h^*_{3,1},...,h^*_{3,n_1})'\in \mathbb{R}^{2n}$. 

Multiplying (\ref{eq:term1}) and (\ref{eq:term2}) together, we have
\begin{equation}
    \begin{aligned} f(\mathbf{z}|\boldsymbol{\xi},\boldsymbol{\beta},\boldsymbol{\eta},\boldsymbol{\theta},\boldsymbol{q})f(\boldsymbol{\xi}|\boldsymbol{\beta},\boldsymbol{\eta},\boldsymbol{\theta},\boldsymbol{q}) \propto \frac{N}{\sigma_{\xi}^n}\exp\left[\boldsymbol{\alpha}_{D,\xi}'\left\{\begin{pmatrix}
\boldsymbol{I}_n & \boldsymbol{X} & \boldsymbol{G} \\
\frac{1}{\sigma_{\xi}}\boldsymbol{I}_n & \boldsymbol{0}_{n,p} & \boldsymbol{0}_{n,3r}
\end{pmatrix}\begin{pmatrix}
\boldsymbol{\xi} \\
\boldsymbol{\beta} \\
\boldsymbol{\eta} 
\end{pmatrix}-\begin{pmatrix}
\boldsymbol{\delta}_y \\
\boldsymbol{\delta}_{\xi}
\end{pmatrix}\right\}\right. \\
\left.-\boldsymbol{\kappa}_{D,\xi}'\boldsymbol{\psi}_{\xi}\left\{\begin{pmatrix}
\boldsymbol{I}_n & \boldsymbol{X} & \boldsymbol{G} \\
\frac{1}{\sigma_{\xi}}\boldsymbol{I}_n & \boldsymbol{0}_{n,p} & \boldsymbol{0}_{n,3r}
\end{pmatrix}\begin{pmatrix}
\boldsymbol{\xi} \\
\boldsymbol{\beta} \\
\boldsymbol{\eta} 
\end{pmatrix}-\begin{pmatrix}
\boldsymbol{\delta}_y \\
\boldsymbol{\delta}_{\xi}
\end{pmatrix}\right\}\right],
        \label{eq:term12}
    \end{aligned}
\end{equation}
where $\boldsymbol{\alpha}_{D,\xi}=(\mathbf{z}_1'\boldsymbol{D}_{\sigma,1}',\mathbf{z}_2'\boldsymbol{D}_{\sigma,2}',\mathbf{z}_3'+\alpha_{\xi}\boldsymbol{1}_{1,n_1},\boldsymbol{0}_{1,n})'$ and $\boldsymbol{\kappa}_{D,\xi}=(\frac{1}{2}\boldsymbol{1}_{1,n_1^*}\boldsymbol{D}_{\sigma,1}',\frac{1}{2}\boldsymbol{1}_{1,n_2}\boldsymbol{D}_{\sigma,2}',\boldsymbol{1}_{1,n_1}+2\alpha_{\xi}\boldsymbol{1}_{1,n_1},\frac{1}{2}\boldsymbol{1}_{1,n})'$. Then we multiply (\ref{eq:term12}) by $f(\boldsymbol{\beta}|\boldsymbol{\theta},\boldsymbol{q})f(\boldsymbol{\eta}|\boldsymbol{\theta},\boldsymbol{q})f(\boldsymbol{q})\pi(\boldsymbol{\theta})$ and obtain
\begin{equation}
    \begin{aligned} f(\boldsymbol{\xi},\boldsymbol{\beta},\boldsymbol{\eta},\boldsymbol{\theta},\boldsymbol{q}|\mathbf{z}) \propto \frac{\pi(\boldsymbol{\theta})N}{\mathrm{det}\{\boldsymbol{D}(\boldsymbol{\theta})\}}\exp\left[\boldsymbol{\alpha}'\left\{\begin{pmatrix}
\boldsymbol{I}_n & \boldsymbol{X} & \boldsymbol{G} \\
\boldsymbol{0}_{p,n} & \boldsymbol{D}_{\beta}(\boldsymbol{\theta})^{-1} & \boldsymbol{0}_{p,3r} \\
\boldsymbol{0}_{3r,n} & \boldsymbol{0}_{3r,p} & \boldsymbol{D}_{\eta}(\boldsymbol{\theta})^{-1} \\
\frac{1}{\sigma_{\xi}}\boldsymbol{I}_n & \boldsymbol{0}_{n,p} & \boldsymbol{0}_{n,3r}
\end{pmatrix}\begin{pmatrix}
\boldsymbol{\xi} \\
\boldsymbol{\beta} \\
\boldsymbol{\eta} 
\end{pmatrix}-\begin{pmatrix}
\boldsymbol{\delta}_y \\
\boldsymbol{\delta}_{\beta} \\
\boldsymbol{\delta}_{\eta} \\
\boldsymbol{\delta}_{\xi}
\end{pmatrix}\right\}\right. \\
\left.-\boldsymbol{\kappa}'\boldsymbol{\psi}\left\{\begin{pmatrix}
\boldsymbol{I}_n & \boldsymbol{X} & \boldsymbol{G} \\
\boldsymbol{0}_{p,n} & \boldsymbol{D}_{\beta}(\boldsymbol{\theta})^{-1} & \boldsymbol{0}_{p,3r} \\
\boldsymbol{0}_{3r,n} & \boldsymbol{0}_{3r,p} & \boldsymbol{D}_{\eta}(\boldsymbol{\theta})^{-1} \\
\frac{1}{\sigma_{\xi}}\boldsymbol{I}_n & \boldsymbol{0}_{n,p} & \boldsymbol{0}_{n,3r}
\end{pmatrix}\begin{pmatrix}
\boldsymbol{\xi} \\
\boldsymbol{\beta} \\
\boldsymbol{\eta} 
\end{pmatrix}-\begin{pmatrix}
\boldsymbol{\delta}_y \\
\boldsymbol{\delta}_{\beta} \\
\boldsymbol{\delta}_{\eta} \\
\boldsymbol{\delta}_{\xi}
\end{pmatrix}\right\}\right],
        \label{eq:termall}
    \end{aligned}
\end{equation}
where $\boldsymbol{\alpha}=(\mathbf{z}_1'\boldsymbol{D}_{\sigma,1}',\mathbf{z}_2'\boldsymbol{D}_{\sigma,2}',\mathbf{z}_3'+\alpha_{\xi}\boldsymbol{1,n_1},\boldsymbol{0}_{1,n+p+3r})'$, $\boldsymbol{\kappa}=(\frac{1}{2}\boldsymbol{1}_{1,n_1^*}\boldsymbol{D}_{\sigma,1}',\frac{1}{2}\boldsymbol{1}_{1,n_2}\boldsymbol{D}_{\sigma,2}',\boldsymbol{1}_{1,n_1}+2\alpha_{\xi}\boldsymbol{1}_{1,n_1},\frac{1}{2}\boldsymbol{1}_{1,n+p+3r})'$, $\boldsymbol{D}(\boldsymbol{\theta})^{-1}=blkdiag(\boldsymbol{I}_n,\boldsymbol{D}_{\beta}(\boldsymbol{\theta})^{-1},\boldsymbol{D}_{\eta}(\boldsymbol{\theta})^{-1},\frac{1}{\sigma_\xi}\boldsymbol{I}_n)$, and the unit-log partition function $\boldsymbol{\psi}(\boldsymbol{h})=(\psi_G(h_{1,1}),...,\psi_G(h_{1,n_1^*}),\psi_G(h_{2,1}),...,\psi_G(h_{2,n_2}),\psi_B(h_{3,1}),...,\psi_B(h_{3,n_1}),$\newline$\psi_G(h^*_{1,1}),...,\psi_G(h^*_{1,n_1^*}),\psi_G(h^*_{2,1}),...,\psi_G(h^*_{2,n_2}),\psi_G(h^*_{3,1}),...,\psi_G(h^*_{3,n_1}),\psi_G(h^*_{\beta,1}),...,\psi_G(h^*_{\beta,p}),$\newline$\psi_G(h^*_{\eta,1}),...,\psi_G(h^*_{\eta,3r}))'$ for any $\boldsymbol{h}=(h_{1,1},...,h_{1,n_1^*},h_{2,1},...,h_{2,n_2},h_{3,1},...,h_{3,n_1},h^*_{1,1},...,h^*_{1,n_1^*},$\newline$h^*_{2,1},...,h^*_{2,n_2},h^*_{3,1},...,h^*_{3,n_1},h^*_{\beta,1},...,h^*_{\beta,p},h^*_{\eta,1},...,h^*_{\eta,3r})'\in \mathbb{R}^{2n+p+3r}$.

Then substituting $\boldsymbol{\delta}=-\boldsymbol{D}(\boldsymbol{\theta})^{-1}\boldsymbol{Qq}$ into (\ref{eq:termall}) and taking the integral over $\boldsymbol{\theta}$ would lead us to
\begin{align*}
f(\boldsymbol{\xi},\boldsymbol{\beta},\boldsymbol{\eta},\boldsymbol{q}|\mathbf{z}) &\propto \int_{\Omega} \frac{\pi(\boldsymbol{\theta})N^*}{\mathrm{det}\{\boldsymbol{D}(\boldsymbol{\theta})\}}  \exp\left[\boldsymbol{\alpha}'\boldsymbol{D}(\boldsymbol{\theta})^{-1}(\boldsymbol{H},\boldsymbol{Q})\begin{pmatrix}
    \boldsymbol{\xi}\\
    \boldsymbol{\beta}\\
    \boldsymbol{\eta}\\
    \boldsymbol{q}
\end{pmatrix}-\boldsymbol{\kappa}'\boldsymbol{\psi}\left\{\boldsymbol{D}(\boldsymbol{\theta})^{-1}(\boldsymbol{H},\boldsymbol{Q})\begin{pmatrix}
    \boldsymbol{\xi}\\
    \boldsymbol{\beta}\\
    \boldsymbol{\eta}\\
    \boldsymbol{q}
\end{pmatrix}\right\}\right]\,\,d\boldsymbol{\theta} \\
&\propto GCM(\boldsymbol{\alpha},\boldsymbol{\kappa},\boldsymbol{0}_{2n+p+3r,1},\boldsymbol{V},\pi,\boldsymbol{D};\boldsymbol{\psi})
\end{align*}
where $N^*=\left(\frac{1}{2\pi}\right)^{(n_1^*+n_2+n+p+3r)/2}\cdot \prod_{i=1}^{n_1^*}\frac{\exp(-Z_1(\boldsymbol{s}_i)^2/2\sigma_{1,i}^2)}{\sigma_{1,i}}\cdot\prod_{j=1}^{n_2}\frac{\exp(-Z_2(A_j)^2/2\sigma_{2,j}^2)}{\sigma_{2,j}}\cdot$\newline$\prod_{k=1}^{n_1}=\frac{\Gamma(1+2\alpha_{\xi})}{\Gamma(Z_3(\boldsymbol{s}_k)+\alpha_{\xi})\Gamma(1-Z_3(\boldsymbol{s}_k)+\alpha_{\xi})}$.

\section{Proof of Equation (\ref{eq:cross:signal:noise})}

In this appendix, we present the proof of Equation (\ref{eq:cross:signal:noise}). We want to show that $cov(\mathbf{y}^*,\boldsymbol{\delta}_y|\boldsymbol{z})$ has the form stated in Equation (\ref{eq:cross:signal:noise}). Note that 
\begin{align*}
    \mathbf{y}^*&=(\boldsymbol{I}_n,\boldsymbol{0}_{n,p},\boldsymbol{0}_{n,3r},\boldsymbol{0}_{n,n})\boldsymbol{H}\boldsymbol{\zeta},\\
    \boldsymbol{\delta}_y&=(\boldsymbol{I}_n,\boldsymbol{0}_{n,p},\boldsymbol{0}_{n,3r},\boldsymbol{0}_{n,n})\boldsymbol{Q}\boldsymbol{q},
\end{align*}
where $\boldsymbol{I}_n$ is an $n\times n$ identity matrix and $\boldsymbol{0}_{i,j}$ is an $i\times j$ zero matrix. We let $\boldsymbol{J}=(\boldsymbol{I}_n,\boldsymbol{0}_{n,p},\boldsymbol{0}_{n,3r},\boldsymbol{0}_{n,n})$. Then from Equation (9) in the main text,
\begin{align*}
    cov(\mathbf{y}^*,\boldsymbol{\delta}_y|\boldsymbol{z},\boldsymbol{\theta}) &= cov\{\boldsymbol{J}\boldsymbol{H}(\boldsymbol{H}'\boldsymbol{H})^{-1}\boldsymbol{H}'\boldsymbol{D}(\boldsymbol{\theta})\boldsymbol{w},\boldsymbol{J}\boldsymbol{Q}\boldsymbol{Q}'\boldsymbol{D}(\boldsymbol{\theta})\boldsymbol{w}|\boldsymbol{z},\boldsymbol{\theta}\}\\
    &= \boldsymbol{J}\boldsymbol{H}(\boldsymbol{H}'\boldsymbol{H})^{-1}\boldsymbol{H}'\boldsymbol{D}(\boldsymbol{\theta})cov(\boldsymbol{w}|\boldsymbol{\alpha},\boldsymbol{\kappa})\boldsymbol{D}(\boldsymbol{\theta})'\boldsymbol{Q}\boldsymbol{Q}'\boldsymbol{J}'\\
    &= \boldsymbol{J}\boldsymbol{H}(\boldsymbol{H}'\boldsymbol{H})^{-1}\boldsymbol{H}'\boldsymbol{D}(\boldsymbol{\theta})cov(\boldsymbol{w}|\boldsymbol{\alpha},\boldsymbol{\kappa})\boldsymbol{D}(\boldsymbol{\theta})'\{\boldsymbol{I}_{2n+p+3r}-\boldsymbol{H}(\boldsymbol{H}'\boldsymbol{H})^{-1}\boldsymbol{H}'\}\boldsymbol{J}',
\end{align*}
which does not equal to zero. Therefore, the posterior cross-signal-to-noise dependence is implied.

 \section{ROC Curve}

 \begin{figure}[!ht]
		\centering
		\includegraphics[width=0.6\textwidth]{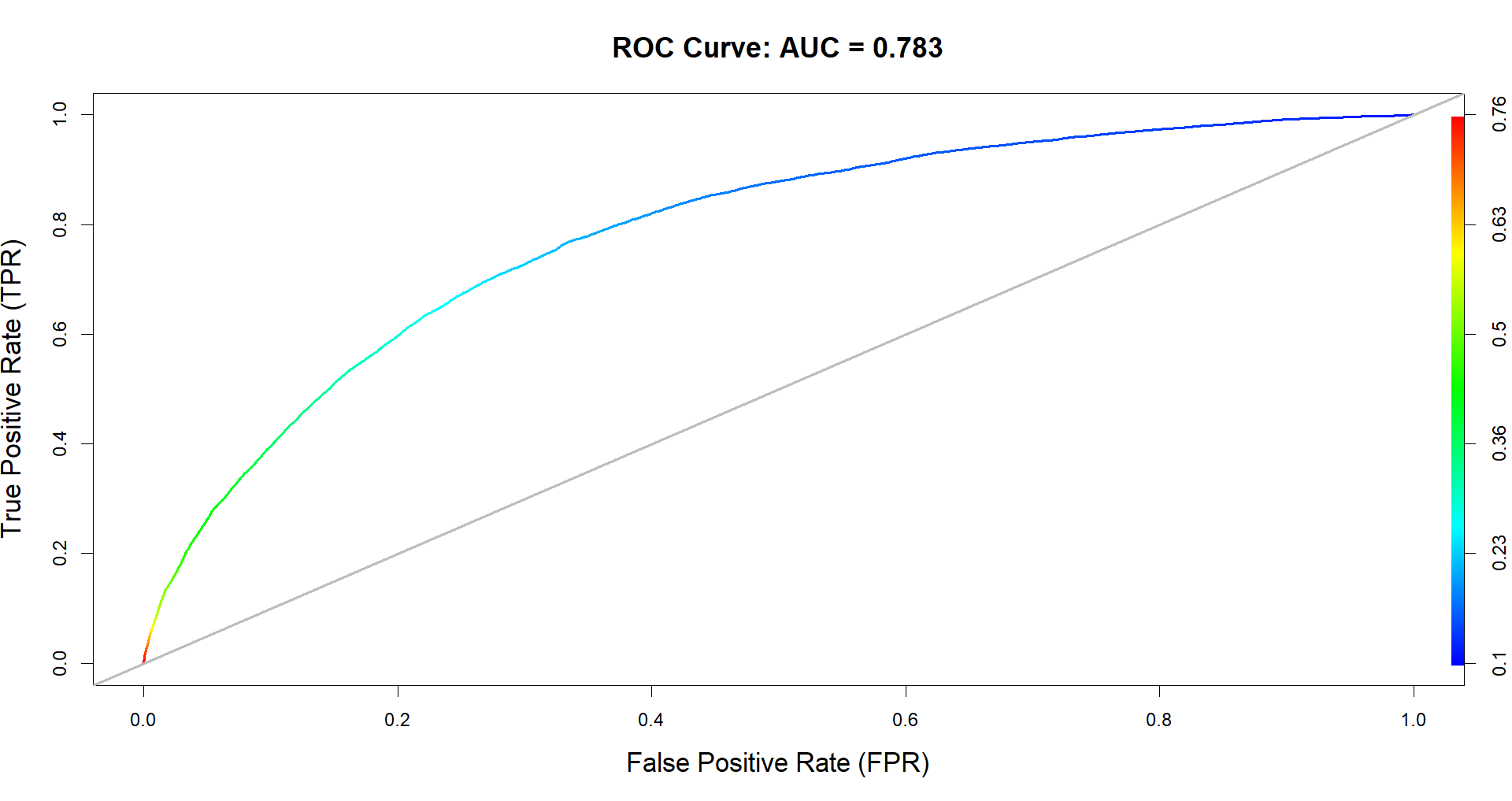} 
	\caption{The Receiver Operating Characteristic (ROC) curve for fire indicator variable. The curve plots the true positive rate against the false positive rate with the color gradient showing various threshold settings. The diagonal grey line from the bottom left to the top right indicates a classifier's performance that is no better than random guessing and serves as a benchmark for comparison. The computed area under the curve (AUC) is reported in the plot title. }
	\label{fig:roc}
\end{figure}

\clearpage
 
\section{Plots of Predictions}
 
    \begin{figure}[!ht]
	\centering
	\begin{subfigure}[t]{0.6\textwidth}
		\centering
		\includegraphics[width=\linewidth]{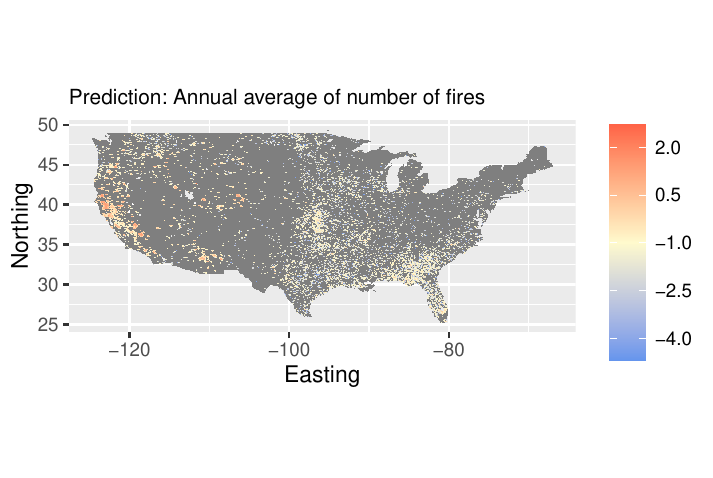} 
		\caption{}
		\label{fig:fire_eprpred}
	\end{subfigure}
	
	\begin{subfigure}[t]{0.6\textwidth}
		\centering
		\includegraphics[width=\linewidth]{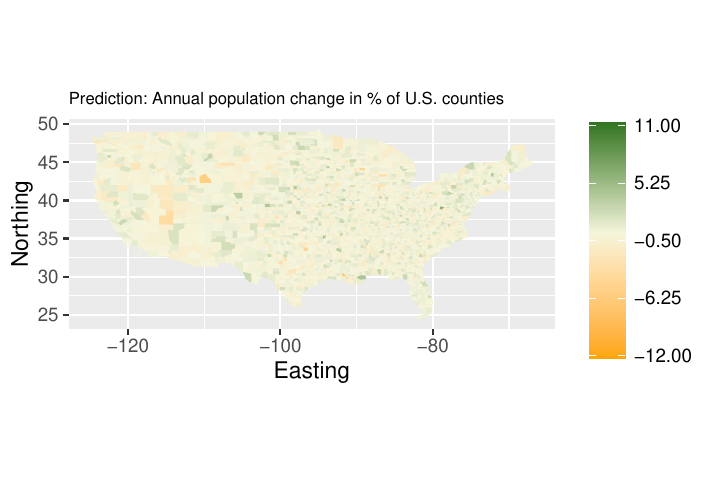} 
		\caption{}
		\label{fig:pop_eprpred}
	\end{subfigure}
	\caption{Predictions of active fires and population changes in the United States from July 2020 to June 2021. (a) Predicted annual average of number of fires over 80,817 point-referenced locations. The values are presented on log scale for illustration purpose only. The grey areas indicate that there are no fires at those locations. (b) Predicted annual population change in percentage over 3,109 counties. Discontinuities are seen as county borders are removed for better illustration.}
	\label{fig:firenpop_pred}
\end{figure}

\clearpage

\section{Gelman-Rubin Statistic}

\begin{table}[!ht]
	\centering
	\begin{tabular}{c@{\hskip 0.2in} c@{\hskip 0.2in} c}
		\hline
		\hline
		Parameter & Point Estimate & Upper C.I. \\
		\hline
		$\boldsymbol{\beta}_1$ & 557 & 1433 \\
		$\boldsymbol{\beta}_2$ & 21.8 & 125 \\
		$\boldsymbol{\beta}_3$ & 951 & 3612 \\
		$\boldsymbol{\eta}_1$ & 29.7 & 184 \\
		$\boldsymbol{\eta}_2$ & 7.54 & 45.8 \\
		$\boldsymbol{\eta}_3$ & 1.29 & 1.39 \\
		$\boldsymbol{\xi}_1$ & 8.78 & 51.4 \\
		$\boldsymbol{\xi}_2$ & 1.31 & 2.24 \\
		$\boldsymbol{\xi}_3$ & 26.1 & 162 \\
		$\sigma^2_1$ & 147 & 901 \\
		$\sigma^2_2$ & 18.6 & 41.5 \\
		$\sigma^2_{\eta}$ & 19 & 118 \\
		$\sigma^2_{\beta}$ & 3.17 & 17.8 \\
		$\sigma^2_{\xi}$ & 377 & 2341 \\
		\hline
		\hline
	\end{tabular}
	\caption{Point estimate and 95\% upper confidence limit of Gelman-Rubin statistic for each parameter computed using 2 chains and 10,000 iterations each with first 5,000 discarded for burn-in. Note that for vector parameters, values are computed for the first element of the vector.}
\end{table}

    \end{appendices}

	\bigskip
	\bigskip

	\clearpage


\begin{thebibliography}{11}
\bibitem[Bradley et al.(2020)]{bradley20} Bradley, J.R., Wikle, C.K. and Holan, S.H. (2020) Hierarchical Models for Spatial Data with Errors that are Correlated with the Latent Process. \textit{Statistica Sinica}, 30, 80--109.
\end{thebibliography}
\end{document}